# *Flexible Dynamic Information Flow Control in the Presence of Exceptions*

Deian Stefan[1] Alejandro Russo[2] John C. Mitchell[1] David Mazières[1]
(1) Stanford University, Stanford, CA, USA
(2) Chalmers University of Technology, Gothenburg, Sweden
(*e-mail:* {deian,mitchell}@cs.stanford.edu, russo@chalmers.se)

**Abstract**

We describe a new, dynamic, floating-label approach to language-based information flow control. A labeled IO monad, `LIO`, keeps track of a *current label* and permits restricted access to IO functionality. The current label floats to exceed the labels of all data observed and restricts what can be modified. Unlike other language-based work, LIO also bounds the current label with a *current clearance* that provides a form of discretionary access control. Computations may encapsulate and pass around the results of computations with different labels. In addition, the `LIO` monad offers a simple form of labeled mutable references and exception handling. We give precise semantics and prove confidentiality and integrity properties of a call-by-name $\lambda$-calculus and provide an implementation in Haskell.

## 1 Introduction

Complex software systems are often composed of modules with different provenance, trustworthiness, and functional requirements. A central security design principle is the *principle of least privilege*, which says that each component should be given only the privileges it needs for its intended purpose. In particular, it is important to differentially regulate access to sensitive data in each section of code. This minimizes the trusted computing base for each overall function of the system and limits the downside risk if any component is either maliciously designed or compromised.

Information flow control (IFC) tracks the flow of sensitive data through a system and prohibits code from operating on data in violation of a security policy. Significant research, development, and experimental effort has been devoted to static information flow mechanisms. Static analysis has a number of benefits, including reduced run-time overhead, fewer run-time failures, and robustness against implicit flows (Denning & Denning, 1977). However, static analysis does not work well in environments where new classes of users and new kinds of data are encountered at run-time. In order to address the needs of such systems, we describe a new, dynamic, floating-label approach to language-based information flow control and present an implementation in Haskell.

Our approach uses a `Labeled` type constructor to protect values by associating them with labels. The labels themselves are typed values manipulated at run-time, and can thus be created dynamically based on other data such as a username. Conceptually, at each point in the computation, the evaluation context has a *current label*. We use a labeled IO monad, `LIO`, to keep track of the current label and permit restricted access to IO functionality (such as a labeled file system), while ensuring that the current label accurately represents an upper bound on the labels of all data observed or modified. Unlike other language-based work, `LIO` also bounds the current label with a *current clearance*. The clearance of a region of code may be set in advance to impose an upper bound on the floating current label within that region. This restricts data access, limits the amount of code that could manipulate sensitive data, and reduces opportunities to exploit covert channels. Additionally, we introduce an operator, `toLabeled`, that allows the result of a computation that would have raised the current label to be encapsulated within the `Labeled` type. Finally, we present combinators for working with labeled references, and exceptions. Thanks to the flexibility of dynamic checking, `LIO` implements an IFC mechanism that is more permissive than previous static approaches (Pottier & Simonet, 2002; Li & Zdancewic, 2010; Russo *et al.*, 2008) but provides similar security guarantees (Sabelfeld & Russo, 2009). Though purely language-based, LIO explores a new design point centered on floating labels that draw on past OS work (Zeldovich *et al.*, 2006).

The main features of our system can be understood using the example of an online conference review system, called $\lambda$Chair. In this system, which we describe more fully later in the paper, authenticated users can read any paper and can normally read any review. This reflects the normal practice in conference reviewing, for example, where every member of the program committee can see submissions and their reviews, and participate in related discussion. Users can be added dynamically and assigned to review specific papers. As an illustration of the power of the labeling system, integrity labels are used to make sure that only assigned reviewers can write reviews for any given paper. Conversely, confidentiality labels are used to manage conflicts of interest. Users with a conflict of interest on a specific paper lack the privileges, represented by confidentiality labels, to read a review. As conflicts of interest are identified, confidentiality labels on the papers may change dynamically and become more restrictive.

This paper extends an earlier conference version (Stefan *et al.*, 2011b) by including formal proofs and extending the calculus and library implementation with exception handling. The main contributions of this work are:

- We propose a new design point for IFC systems in which most values in lexical scope are protected by a single, mutable, *current label*, yet one can also encapsulate and pass around the results of computations with different labels. Label encapsulation is explicitly reflected by types in a way that prevents implicit flows.
- We prove information flow and integrity properties of our design and describe LIO, an implementation of the new model in Haskell. LIO, which can be implemented entirely as a library (relying solely on type safety), demonstrates both the applicability and simplicity of the approach.
- Unlike other language-based work, our model provides a notion of *clearance* that imposes an upper bound on the program label, thus providing a form of discretionary access control on portions of the code, i.e., restricting access to data it "needs to know".
- We present a novel dynamic, yet safe, handling of exceptions. Exceptions are a key component to make LIO a more practical IFC system.

This paper is organized as follows. Section 2 provides background on information flow control and our Haskell LIO library. Section 3 presents a motivating scenario where to apply LIO. Formalization of the library is given in Section 4 and the security guarantees are detailed in Section 5. Related work is described in Section 6. We conclude in Section 7.

## 2 Security Library

In this section, we give an overview of information flow control, the approach used by LIO to dynamically enforce IFC, and the core application programming interface (API) provided by our library.

**Labels and IFC** The goal of information flow control is to track the propagation of information and control it according to a security policy. A well-known policy addressed in almost every IFC system is *non-interference:* publicly-readable program results must not depend on secret inputs. A non-interfering program is guaranteed to preserve confidentiality of sensitive data (Goguen & Meseguer, 1982); dually, this policy can be used to preserve integrity of trustworthy data (Biba, 1977).

To enforce information flow restrictions, most systems associate *labels* with every piece of data. A label represents the level of confidentiality and integrity on data. Labels form a lattice (Denning, 1976) with partial order $\sqsubseteq$ (pronounced "can flow to"); $\sqsubseteq$ is used to govern the allowed flows between differently labeled entities. For instance, if $L_1 \sqsubseteq L_2$ holds, it indicates that data with label $L_1$ can flow into entities labeled $L_2$.

LIO is polymorphic in the label type, allowing different types of labels to be used. Custom label formats can be created by providing a definition for a bounded lattice. Specifically, a label format must have a well-defined partial order ($\sqsubseteq$), a binary operation computing the *join* of two labels ($\sqcup$), a binary operation computing the *meet* of two labels ($\sqcap$), and minimum ($\bot$) and maximum ($\top$) elements. For any two labels $L_1$ and $L_2$, the join has the property that $L_i \sqsubseteq (L_1 \sqcup L_2), i = 1,2$ and $L_1 \sqcup L_2$ is the least of such elements in the lattice; the meet has the property that $(L_1 \sqcap L_2) \sqsubseteq L_i, i = 1,2$ and $L_1 \sqcap L_2$ is the greatest of such elements in the lattice. In our Haskell library, label types are instances of the `Label` type class:

```
class (Eq l) ⇒ Label l where
  leq  :: l → l → Bool -- Can flow to relation (⊑)
  lub  :: l → l → l    -- Join operation (⊔)
  glb  :: l → l → l    -- Meet operation (⊓)
  lbot :: l            -- Minimum element (⊥)
  ltop :: l            -- Maximum element (⊤)
```

Henceforth we assume that the bounded lattice property holds for the labels used in our examples. Section 3 details disjunction category (DC) labels, a concrete label format used int $\lambda$Chair that satisfies this property.

**Privileges and Decentralized IFC** An extension of IFC, the decentralized label model (DLM) of Myers and Liskov (Myers & Liskov, 1997) allows for more general applications, including systems consisting of mutually distrustful parties. In a decentralized system, a computation is executed with a set of *privileges* $p$, which, when exercised, allow the computation to "bypass" certain label restrictions. In such systems, rather than using the standard $\sqsubseteq$ partial order relation, a more permissive pre-order $\sqsubseteq_p$, is used in the label comparisons. Consider, for example, a simple four-point lattice $\bot \sqsubseteq L_i \sqsubseteq L_{AB}$, for $i = A, B$. Here, $L_A$, $L_B$, and $L_{AB}$ respectively correspond to data private to user $A$, user $B$, and both $A$ and $B$. In DLM, privileges and labels are associated such that, e.g., privilege $a$ corresponding to $L_A$ allows user $A$ to "ignore the $A$" in labels. Thus, $L_{AB} \sqsubseteq_a L_B$, even though $L_{AB} \not\sqsubseteq L_B$. This property is very useful as it allows $A$ to *downgrade* the data from label level $L_{AB}$ to $L_B$. Informally, when downgrading, the code exercising the privilege states that it no longer considers the data to be confidential (in this case, $A$ exercising $a$ to downgrade data from $L_{AB}$ to $L_B$). Note that downgrading does not make the data publicly readable, all parties corresponding to the label must first perform the downgrade.

As in the case of labels, our library is polymorphic in the privilege types. Any code can exercise privileges (that are in lexical scope) to enforce IFC using the more permissive $\sqsubseteq_p$ relation. However, our formalism is limited to non-privileged primitives and we thus do not discuss privileges further. We refer the interested reader to the library documentation for details on privileges.

**LIO computations** LIO is a language-based *floating-label* system, inspired by IFC operating systems, including HiStar (Zeldovich *et al.*, 2006) and Asbestos (Efstathopoulos *et al.*, 2005). In a floating-label system, the label of a computation can rise to accommodate reading sensitive data, similar to the *program counter* of more traditional language-based systems (Sabelfeld & Myers, 2003). Specifically, in LIO, a computation $C$ with label $L_C$ wishing to observe an object (e.g., a review) labeled $L_R$ can do so by first raising its label to the join of the labels: $L_C \sqcup L_R$. Consider, for example, a simple $\lambda$Chair review system computation that retrieves the content of a review, and writes it to an output channel.

```
readReview R = do            -- Initial computation label: L_C
  rv ← retrieveReview R      -- Computation label when retrieving: L_C ⊔ L_R
  printLabeledCh rv          -- Computation label when printing: L_C ⊔ L_R
```

Here, we assume that the computation is executing on behalf of a user, Clarice, with initial label $L_C$ and that review $R$ has label $L_R$. The computation label is shown in the comments as the different actions are executed. Internally, the `retrieveReview` function is used to retrieve the review contents `rv`; the function raises the computation label to $L_C \sqcup L_R$ to reflect the observation of sensitive review information. This directly highlights the notion of a "floating-label": a computation's label effectively "floats above" the labels of all objects it observes.

The floating label is used to restrict writes: a computation cannot write to an entity whose label is below the computation label. In the example, the action `printLabeledCh rv`, which performs a write, does not change the computation label. However, `printLabeledCh` returns an action that writes the review content `rv` to standard output channel, labeled $L_O$, only if $L_C \sqcup L_R \sqsubseteq L_O$. In $\lambda$Chair, the standard output channel label, $L_O$, is dynamically set according to the user executing the computation; $L_O$ is set so as to allow for printing out all but the conflicting reviews. Thus, if user Clarice has a conflict of interest with review $R$, $L_O$ is set such that $L_R \not\sqsubseteq L_O$.

Unlike existing language-based IFC systems, LIO also associates a *clearance* with each computation. This clearance sets an upper bound on the current floating label within some region of code. For example, the notion of clearance can be used to prevent Clarice from retrieving (and not just printing) the contents of a conflicting review $R$ by setting the computation's clearance to $C_C$ such that $L_R \not\sqsubseteq C_C$. In general, before raising the computation label, LIO combinators first check that the new label will not exceed the computation clearance. Hence, when the action `retrieveReview` $R$ attempts to raise the current label to $L_C \sqcup L_R$, the computation will fail since $L_C \sqcup L_R \not\sqsubseteq C_C$.

More interestingly, clearance can be used to prevent malicious code from exploiting covert channels. For example, without clearance, the following function can be used by a user, such as Clarice, to leak information on reviews which she is in conflict with:

```
leakingRetriveReview r = do -- Initial label: L_C
  rv ← retrieveReview r     -- Retrieving: L_C ⊔ L_R (if L_C ⊔ L_R ⊑ C_C)
  covertChannel rv          -- Leak review into covert channel
```

The function `covertChannel` leaks (part of) the sensitive review content into a covert channel, such as the termination channel. In the latter case, the function leaks information by deciding whether or not to diverge based on sensitive data. A simple example that leaks a bit is given below.

```
covertChannel rv =
  if rv=="Paper..."           -- If sensitive review matches "Paper..."
    then forever (return rv)  -- then loop forever
    else return rv            -- otherwise return
```

Using clearance, we prevent such leaks by setting the clearance and review labels such that `retrieveReview` fails when it attempts to raise the computation label to retrieve conflicting reviews (the additional check $L_C \sqcup L_R \sqsubseteq C_C$ will not hold).

### *2.1 Library Interface*

LIO is a *termination-insensitive* and *flow-sensitive* (Askarov *et al.*, 2008; Hunt & Sands, 2006) IFC library that *dynamically* enforces information flow restrictions. At a high level, LIO defines a monad called `LIO`, intended to be used in place of `IO`. The library furthermore contains a collection of `LIO` actions, many of them similar to `IO` actions from standard Haskell libraries, except that they contain label checks that enforce IFC.

To implement the notion of floating label that is bounded by a clearance, our library defines `LIO` as a state monad, parametric in the label type, and using `IO` as the underlying base monad. The state consists of a *current label* $L_{\text{cur}}$, i.e., the computation's floating label, and a *current clearance* $C_{\text{cur}}$, which is an upper bound on $L_{\text{cur}}$, i.e., $L_{\text{cur}} \sqsubseteq C_{\text{cur}}$ always holds. The (slightly simplified) `LIO` monad is defined as:

```
newtype LIO l a = LIOTCB (StateT (l, l) IO a)
```

where the state corresponds to the current label and clearance. To allow for the execution of `LIO` actions, our library provides the function `evalLIO` that takes an `LIO` action and returns an `IO` action which, when executed, will return the result of the IFC-respecting computation. It is important to note that untrusted LIO code cannot execute `IO` computations by binding `IO` actions with `LIO` ones (to bypass IFC restrictions), because `LIOTCB` is a private symbol. Effectively this limits `evalLIO` to trusted code. Additionally, using `evalLIO`, trusted programmers can easily, though cautiously, enforce IFC in parts of an otherwise IFC-unaware program.

The current label provides a means for associating a label with every piece of data. Hence, rather than individually labeling definitions and bindings, all symbols in scope are protected by $L_{\text{cur}}$. Moreover, the only way to read or modify differently labeled data is to execute (trusted) actions that internally access restricted symbols and appropriately validate and adjust the current label (or clearance).

In many practical situations, it is essential to be able to manipulate differently-labeled data without monotonically increasing the current label. For this purpose, the library additionally provides a `Labeled` type for labeling values with labels other than $L_{\text{cur}}$. A `Labeled`, polymorphic in the label type, protects an immutable value with a specified label irrespective of the current label. This is particularly useful as it allows a computation to delay raising its current label until necessary. For example, an alternative `retrieveReview` implementation can retrieve the review content, convert it to HTML, encapsulate the markup into a `Labeled` value, and return the `Labeled` value while leaving the current label unchanged. This approach delays the "creeping" of current label until the review content, as encapsulated by `Labeled`, is actually *needed*.

We note that `LIO` can be used to protect pure values in a similar fashion to `Labeled`. However, the protection provided by `Labeled` allows for serializing labeled values and straight forward inspection by trusted code (which may ignore the protecting label). Unlike `LIO`, `Labeled` is not a monad[1]. The monad instance would allow a computation to use bind and return to arbitrarily manipulate labeled values without any notion of the current label or clearance, and thus (possibly) violate the restriction that `LIO` computations should not handle values below their current label or above their current clearance. Moreover, the `Monad` instance would require a definition for a default label necessary when lifting a value with `return`. Instead, our library provides several functions that allow for the creation and usage of labeled values within `LIO`. Specifically, we provide (among other) the following functions:

- `label :: Label l ⇒ l → a → LIO l (Labeled l a)`
  Given a label $l$ such that $L_{\text{cur}} \sqsubseteq l \sqsubseteq C_{\text{cur}}$ and a value $v$, the action `label` $l$ $v$ returns a labeled value that protects $v$ with $l$.
- `unlabel :: Label l ⇒ Labeled l a → LIO l a`
  Assuming that `lv` is associated to label $l$, the action `unlabel lv` raises the current label to $L_{\text{cur}} \sqcup l$ if $L_{\text{cur}} \sqcup l \sqsubseteq C_{\text{cur}}$ and returns the unlabeled value. Note that the new current label is at least as high as `lv`'s label, preserving the confidentiality of the value.
- `toLabeled :: Label l ⇒ l → LIO l a → LIO l (Labeled l a)`
  Given a label $l$ such that $L_{\text{cur}} \sqsubseteq l \sqsubseteq C_{\text{cur}}$ and an `LIO` action $m$, `toLabeled` $l$ $m$ executes $m$ without raising the current label. However, instead of returning the result directly, the function returns the result of $m$ encapsulated in a `Labeled`. The label of the labeled value is $l$; to preserve confidentiality (see Section 4 for further details), action $m$ must not read any values with a label above $l$. In monadic terms, `toLabeled` is an environment-oriented action that provides a different context for a temporary bind thread.
- `labelOf :: Label l ⇒ Labeled l a → l`
  If `lv` is a labeled value with label $l$ and value $v$, `labelOf lv` returns $l$.

[1] In fact, `Labeled` cannot be a functor; this would violate non-interference when considering integrity into the security labels.

Our library additionally provides labeled alternatives to mutable references, i.e., `IORef`s. Specifically, we provide labeled references `LIORef l a` that are created with `newLIORef`, read with `readLIORef`, and written to with `writeLIORef`. When creating or writing to a reference with label $L_R$, it must be the case that $L_{\text{cur}} \sqsubseteq L_R \sqsubseteq C_{\text{cur}}$; when reading $L_{\text{cur}}$ is raised to $L_{\text{cur}} \sqcup L_R$, clearance permitting.

In the conference version of this work (Stefan *et al.*, 2011b), the execution of programs stop when the IFC constraints imposed by the $\sqsubseteq$-relationship are not fulfilled. Similar to other dynamic IFC approaches (e.g., (Askarov & Sabelfeld, 2009; Russo & Sabelfeld, 2009; Austin & Flanagan, 2010)), this design decision restricts the possibilities for programs to recover from failures. Later in this section, we show how to extend LIO with exception handling so that programs can recover from failures or insecure actions without compromising confidentiality or integrity of data.

The formal semantics for the functions described above are given in Section 4; in this section, we illustrate their functionality and use through examples. Specifically, consider the previous example of `readReview`. The internal function `retrieveReview` takes a review identifier $R$ and returns the review contents. Internally, `retrieveReview` must have access to a list of reviews, which are individually protected by different labels. In this model, adding a new review to the system can be implemented as:

```
addReview R L_R rv = do
  r ← label L_R rv          -- Checks L_cur ⊑ L_R ⊑ C_cur
  addToReviewList R r       -- Appends labeled review to internal list
```

where the `addToReviewList` simply adds the `Labeled` review to the internal list. The implementation of `retrieveReview` is similar:

```
retrieveReview R = do          -- Initial label, L_cur = L_C
  r ← getFromReviewList R      -- Retrieving a labeled result, L_cur = L_C
  rv ← unlabel r               -- Unlabel result, raises label to L_cur = L_C ⊔ L_R
  return rv                    -- Returning unlabeled content, L_cur = L_C ⊔ L_R
```

where the `getFromReviewList` retrieves the `Labeled` review from the internal list and `unlabel` removes the protecting label, raising the current label to reflect the observation.

We previously alluded to an alternative implementation of `retrieveReview` that returns the labeled, review content in HTML form while keeping the current label the same. This implementation can be directly leverage the above `retrieveReview`:

```
retrieveReviewHtml R = do            -- Outer: Initial label, L_cur = L_C
  r ← toLabeled (L_C ⊔ L_R) $ do      -- Inner: Initial label, L_cur = L_C
        rv ← retrieveReview R        -- Inner: Retrieve review, L_cur = L_C ⊔ L_R
        return (toHtml rv)           -- Inner: Return review, L_cur = L_C ⊔ L_R
  return r                           -- Outer: Return labeled review, L_cur = L_C
```

Note that although the current label within the inner computation is raised, the outer computation's label does not change—instead the marked-up review content is protected by the label $L_C \sqcup L_R$. Hence, only when the review content is actually needed, `unlabel` can be used to retrieve the content and raise the computation's label accordingly:

```
readReviewHtml R = do                -- Initial label L_cur = L_C
   r ← retrieveReviewHtml R          -- Retrieve labeled review, L_cur = L_C
   -- Perform other computations, such that L_cur = L'_C
   rv ← unlabel r                    -- Unlabel labeled review, L_cur = L'_C ⊔ L_R
   printLabeledCh rv                 -- Print review content, L_cur = L'_C ⊔ L_R
```

### 2.1.1 *Exception handling*

Exception handling is common in real-world applications, and, as already noted, LIO provides support for such constructs. Throwing an exception depends on the information present in the lexical scope. Consequently, LIO labels an exception with the current label ($L_{\text{cur}}$) at the point where the exception is thrown. Specifically, the primitive

```
throwLIO :: (Exception e) ⇒ e → LIO l a
```

takes an arbitrary exception and wraps it into a labeled exception type:

```
data LabeledException l e = ...
```

which itself is an instance of `Exception`. The label of the exception is set to the current label $L_{\text{cur}}$.

Conversely, the primitive:

```
catch :: (Exception e) ⇒ LIO l a → (e → LIO l a) → LIO l a
```

can be used to execute an `LIO` action, using an exception handler to address the case when the computation raises an exception. Suppose the current label and clearance are $L_{\text{cur}}$ and $C_{\text{cur}}$, respectively. Given a computation `m`, and an exception handler `he`, `catch m he` executes `m` and then:

1. if no exception is thrown, the result produced by `catch` is simply the result of `m`, leaving the current label and clearance unchanged (as of the execution of `m`).
2. if an exception with label $l \sqsubseteq C_{\text{cur}}$ is thrown when executing `m`, the current label raised to $L_{\text{cur}} \sqcup l$ and the exception handler is invoked (if the exception type matches). Raising the current label to $L_{\text{cur}} \sqcup l$ before executing the exception handler indicates that the handler must not produce side-effects at security levels lower than the one indicated by the label of the exception.

3. if an exception with label $l \not\sqsubseteq C_{\text{cur}}$ is thrown, the exception label is raised to $L_{\text{cur}} \sqcup l$ and re-thrown (propagated to an outer `catch`).

It is worth remarking that primitive `catch` is the only means for inspecting information related to an exception (e.g., kind of exception, security label, etc.).

**Safe propagation of exceptions** In LIO, the standard *propagation of exceptions up the call stack until reaching the nearest enclosing `catch`* can be used to leak information. Consider the following function:

```
condThrow :: LIORef l Bool → LIO l ()
condThrow secRef = do
  sec ← readLIORef secRef
  if sec then throwLIO ...  else return ()
```

Assuming that `condThrow` is invoked with the current label $L_{\text{cur}}$ and `secRef` has label $L_R$, `throwLIO` raises an exception labeled $L_{\text{cur}} \sqcup L_R$ if the secret value stored in the reference is `True`. The exception label indicates that the exception was raised after performing a secret read.

Although `condThrow` cannot directly be used to leak information, it is important to highlight that the function throws an exception if the secret is `True`, and returns `()` otherwise. Hence, in the presence of `toLabeled`, which restores the current label, it is important to reason about the propagation of exceptions. More specifically, if exceptions propagate until reaching the nearest enclosing `catch`, the following function can be used to leak information:

```
leakIntoPub :: LIORef l Bool → LIORef l Bool → LIO l ()
leakIntoPub secRef pubRef  =  catch (
 do writeLIORef pubRef True              -- Write to public reference #1
    _ ← toLabeled ⊤ $ condThrow secRef -- Throw exception if secret is True
    writeLIORef pubRef False             -- Write to public reference #2
 ) (λ_ → return ())                      -- Handle exception
```

Suppose that the function is invoked with a current label $L_{\text{cur}} = \bot$ and current clearance $C_{\text{cur}} = \top$, `secRef` is labeled $\top$ and `pubRef` is labeled $\bot$. Initially, the computation can directly read and write to `pubRef`, but only read from `secRef`.

Note that `catch` is only used to force normal termination, i.e., execution of function `leakIntoPub` always return `()`. More importantly, note that public side-effects are performed before (`writeLIORef pubRef True`) and after (`writeLIORef pubRef False`) executing a computation on secret data (`condThrow secRef`). (This is possible because the computation `condThrow secRef` is enclosed in a `toLabeled` block, and thus the current label remains unchanged.) Moreover, if the value of the secret reference `secRef` is `True`, then an exception is raised in `condThrow` and further propagated to the enclosing `catch` without executing the second write to the public reference (`writeLIORef pubRef False`). Hence, if an exception is raised in `condThrow` the content of `pubRef` remains `True`. In contrast, if no exception is thrown, the content of `pubRef` is set to `False`: clearly, a direct leak of the value stored in `secRef`.

It is important to finally note that although catch will raise the current label when an exception raised in the secret computation, `leakIntoPub` can also be enclosed by `toLabeled`:

```
leakSecretRef :: LIORef l Bool → LIO l Bool
leakSecretRef secRef = do
  pubRef ← newLIORef ⊥ True                    -- Create public reference
  toLabeled ⊤ $ leakIntoPub secRef pubRef   -- Perform attack
  readLIORef pubRef                            -- Read "secRef" value
```

This function returns the content of the secret reference `secRef` without raising the current label.

Due to the feasibility of such attacks, LIO propagates exceptions up to the nearest `catch` *or* `toLabeled`. Intuitively, the correct semantics of `toLabeled` are as before with the added requirement that all exceptions be caught: regardless how the computation enclosed by `toLabeled` terminates, a `Labeled` value must always be returned. Conceptually this is equivalent to labeling a lifted value, i.e., a value that may be a "normal" value or an exception. Of course, if the result of `unlabel` is an exception, the exception will propagate to the nearest `catch` or `toLabeled`.

Considering this modification to the semantics of `toLabeled`, observe that the side-effects in `leakIntoPub` produced after the `toLabeled` block will always be executed (even if an exception is raised inside `condThrow`). More generally, we close up leaks through exception propagation by simply assuring that the execution of (possibly public) actions following a `toLabeled` block does not depend on the abnormal termination of a computation inside `toLabeled`.

**Recovery of unsafe actions** Unlike other dynamic IFC approaches, such as (Askarov & Sabelfeld, 2009; Sabelfeld & Russo, 2009; Austin & Flanagan, 2009; Austin & Flanagan, 2010; Devriese & Piessens, 2011)), LIO allows untrusted programs to safely recover from failures due to IFC violation attempts (e.g., trying to create labeled values below the current label, or read from a reference labeled above the current clearance, etc.) Having a safe handling of exceptions in place, LIO raises a labeled exception when a security constraint is not fulfilled. This allows untrusted code to catch exceptions and handle monitor failures gracefully. Consider, for instance, the following function that unlabels a `Labeled` value and returns a `Maybe` value to indicate the success of such operation:

```
safeUnlabel :: Labeled a → LIO l (Maybe a)
safeUnlabel lv = catch ( do v ← unlabel lv -- Fails if labelOf lv ⋢ C_cur
                            return (Just v)
                       ) (λ_ → return Nothing)
```

If the label of `lv` is above the current clearance, the LIO primitive `unlabel` throws an exception. In this example, however, this exception is caught (since the label of the exception will be $L_{cur}$ and the exception handler will simply return `Nothing`). If the label of `lv` is below the clearance, the current label is raised and the unlabeled result is simply returned.

## 3 λChair

To demonstrate the flexibility of our dynamic information flow library, we present λChair, a simple API (built on the examples of Section 2) for implementing secure conference reviewing systems. In general, a conference reviewing system should support various features (and security policies) that a program committee can use in the review process. Minimally, it should support:

- *Paper submission*: ability to add new papers to the system.
- *User creation*: ability to dynamically add new reviewers.
- *User login*: a means for authenticating users.
- *Review delegation*: ability to assign reviewers to papers.
- *Paper reading*: means for reading papers.
- *Review writing*: means for writing reviews.
- *Review reading*: means for reading reviews.
- *Conflict establishment*: ability to restrict specific users from reading conflicting reviews.

Even for such a minimal system, a number of security concerns must be addressed. First, only users assigned to a paper may write the corresponding reviews. Second, information from the review of one paper should not leak into a different paper's review. And, third, users should not receive any information on the reviews of the papers with which they are in conflict.

$\lambda$Chair's API provides the aforementioned security policies by applying information flow control. Following the examples of Section 2, we take the approach of enforcing IFC when writing to output channels, and thus the security for the above policies correspond to that of non-interference, i.e., secret data is not leaked into less secret channels/reviews. The alternative, clearance-restricting approach of Section 2 can be used to enforce the security policies by confinement rather than non-interference (see Section 5). Before delving into the details of the $\lambda$Chair, we first introduce the specific label format used in the implementation.

### *3.1 DC Labels*

$\lambda$Chair is implemented using *Disjunction-Category (DC) labels* (Stefan *et al.*, 2011a). DC labels can be used to express a conjunction of restrictions on information flow that represents the interests of multiple stake-holders. As a result, DC labels are especially suitable for systems in which participating parties do not fully trust each other, e.g., a conference review system.

Policies are expressed by leveraging the notions of *principals*. In our system, a principal is a string that represents a source of authority such as a user, group, role, etc. A DC label, written $\langle S, I \rangle$, consists of two Boolean formulas $S$ and $I$ over principals. Both *components* $S$ and $I$ are minimal formulas in conjunctive normal form (CNF), with positive terms and clauses sorted to give each formula a unique representation. Component $S$ protects secrecy by specifying the principals that are allowed (or whose consent is needed) to observe the data. Dually, $I$ protects integrity by specifying principals who created, vouches for, and may currently modify the data.

Data may flow between differently labeled entities, but only in such a way as to accumulate additional secrecy restrictions or be stripped in integrity ones, not vice versa. Specifically, the $\sqsubseteq$-relation for DC labels is defined as:

*Definition 1* (*DC label $\sqsubseteq$ relation*)
For any two DC labels $\langle S_1, I_1 \rangle$ and $\langle S_2, I_2 \rangle$,

$$\frac{S_2 \implies S_1 \qquad I_1 \implies I_2}{\langle S_1, I_1 \rangle \sqsubseteq \langle S_2, I_2 \rangle}$$

In other words, data labeled $\langle S_1, I_1 \rangle$ can flow to an entity labeled $\langle S_2, I_2 \rangle$ if and only if the secrecy of the data, and integrity of the entity are preserved. Intuitively, the $\sqsubseteq$ relation imposes the restriction that any set of principals who can observe data afterwards must also have been able to observe it earlier. Dually, the integrity of the entity is preserved by requiring that the source label impose more restrictions than that of the destination.

The join and meet for DC labels directly follows from the definition. The join and meet of any two DC labels $L_1 = \langle S_1, I_1 \rangle$ and $L_2 = \langle S_2, I_2 \rangle$ are respectively: $L_1 \sqcup L_2 = \langle S_1 \wedge S_2, I_1 \vee I_2 \rangle$ and $L_1 \sqcap L_2 = \langle S_1 \vee S_2, I_1 \wedge I_2 \rangle$, where each component of the resulting labels is reduced to CNF.

Intuitively, the secrecy component of the join protects the secrecy of $L_1$ and $L_2$ by specifying that both set of principals, those appearing in $S_1$ and those in $S_2$, must consent for data labeled $S_1 \wedge S_2$ to be observed. Conversely, the integrity component of the join, $I_1 \vee I_2$, specifies that either principals of $I_1$ or $I_2$ could have created and modify the data. Dual properties hold for the meet $L_1 \sqcap L_2$.

We note that our implementation of DC labels forms a bounded lattice. The least restrictive component corresponds to the Boolean value **True**; the most restrictive component corresponds to the Boolean value **False**. These interpretations allow for a sound definition of the top $\top$ and bottom $\bot$ elements for the DC label lattice: $\top = \langle \textbf{False}, \textbf{True} \rangle$, and $\bot = \langle \textbf{True}, \textbf{False} \rangle$. Additionally, in our model, public entities have the default, or *empty* label, $L_{\text{pub}} = \langle \textbf{True}, \textbf{True} \rangle$. It is intuitive that data labeled $\langle S, I \rangle$ can be written to a public network with label $L_{\text{pub}}$, only with the permission of a set of principals satisfying the Boolean formula $S$. Conversely, data read from the network can be labeled $\langle S, I \rangle$ only with the permission of a set of principals satisfying $I$.

### *3.2 DC Labels in λChair*

We now describe the data structures and the role of DC labels (from now on just labels) in λChair. Intuitively, the λChair API provides administrators and reviewers with functions for querying review entries and modifying user accounts. Hence, λChair  is implemented as a state monad `RevLIO` (whose value constrctor `RevLIOTCB` is not exported to untrusted code) that stores information on reviews and users, with `LIO` as the underlying monad.

The λChairsystem relies on two principal types corresponding to papers and reviews. We identify papers and reviews according to the unique paper identifier/number. As such, for the $i$th paper the principal associated with the paper is $P_i$, while the principal associated with the corresponding review is $R_i$.

**Review entries** A review entry is defined as a record consisting of a paper number, a reference to the corresponding paper, and a reference to the shared review 'notebook'. For simplicity, all reviewers append their review to the same review notebook. The type for such entries is:

```
data ReviewEnt = ReviewEnt { paperId :: Id             -- Paper number
                           , paper   :: DCRef Paper    -- Paper content
                           , review  :: DCRef Review } -- Notebook
```

where `DCRef` is a labeled reference using DC labels, i.e., `type DCRef = LIORef DCLabel`. Note that this differs from the examples of Section 2, where the reviews were simply `Labeled` values.

**Users** A reviewer, or user, has a unique user name, password, and two disjoint sets of paper ids (in our implementation these are simple lists). One set corresponds to the user's conflicting papers, the second set corresponds to the papers the user has been assigned to review. Concretely, we define a user as value of type:

```
data User = User { name        :: Name     -- User name
                 , password    :: Password -- Password
                 , conflicts   :: [Id]     -- Conflicting papers
                 , assignments :: [Id] }   -- Paper assigned to review
```

A user is authenticated given a user name and password as credentials. Following authentication, the code of the reviewer, who is assigned to papers $1, \ldots, n$, is executed with the current label initially set to $\langle \textbf{True},\ R_1 \wedge \cdots \wedge R_n \rangle$, where $R_i$ is the principal corresponding to review entry $i$. The current clearance is set to $\top = \langle \textbf{False},\ \textbf{True} \rangle$. The secrecy component in the clearance allows the executing code to read any data; the integrity component of the current label allow the process to only write to assigned reviews (detailed below).

**Reading and writing papers** After logging in, users are allowed to read and print out any paper by supplying the paper id. The label of the reference `paper` in the $i$th review entry is set to $\langle \textbf{True}, P_i \rangle$. The secrecy component does not restrict any computation from observing the paper by reading the reference content (the paper). However, the integrity component restricts the modification of the $i$th paper to computations that own principal $P_i$ and can therefore run with $P_i$ in the integrity component of its current label. Only a trusted administrator and the paper submission code is allowed to own such principals. As a consequence, computations executing on behalf of a reviewer cannot modify the paper since the current label assigned by the trusted login procedure never includes $P_i$ in its integrity component.

**Reading and writing reviews** A reviewer's code is also allowed to access the review notebook content of arbitrary review entries. Once a review has been read, however, its content must not be leaked into another paper's review notebook. We fulfill this requirement by identifying, using labels, when a given piece of code reads a certain review. Concretely, we label the reference `review` of the $i$th review entry as $\langle R_i, R_i \rangle$. As a consequence, when a computation wishes to read the review for entry $i$, it must raise[2] it current label so as to include the principal $R_i$ in its components (clearance permitted). Once a computation has been *tainted* as such, it will not be able to modify the contents of another paper's `review`. Such tainted computations will have a label with principal $R_i$ in the secrecy component (as a conjunction) and integrity component (as a disjunction). Consider, for instance, a computation performing a review of paper $i$ such that the current label is $L_i = \langle R_i, R_i \rangle$. If the computation subsequently reads a different review labeled $L_j = \left\langle R_j, R_j \right\rangle$, the current label is set to $L = \left\langle R_i \wedge R_j, R_i \vee R_j \right\rangle$. To write to the either review $i$ or $j$ it must be that the current label flows to the review labels, i.e., $L \sqsubseteq L_i$ or $L \sqsubseteq L_j$. It is clear that neither flow restrictions are satisfied, and thus such illegal writes are prevented.

**Conflicts** Following the `readReview` examples of Section 2, we restrict the reading, or more specifically, printing of a review by those reviewers in conflict with the paper. Although every user is allowed to retrieve a review, they cannot observe the result unless they write it to an output channel. Hence, code running on behalf of a user (determined after logging in) can only write to the output channel (using `printLabelCh`) if the current label $L$ can flow to the output channel label $L_o$. Using the set of conflicting paper ids, for every user, we dynamically assign the output channel label $L_o = \langle S_o, \textbf{True} \rangle$, where $S_o = R_1 \wedge \cdots \wedge R_n \wedge (R_{n+1} \vee \texttt{\#CONFLICT}) \wedge \cdots \wedge (R_N \vee \texttt{\#CONFLICT})$ and $R_i$, $i = n+1, \ldots, N$ are the principals corresponding to *all* the review entries in the system (at the point of the print) that the authenticated user is in conflict with. Here, `#CONFLICT` corresponds to a principal that none of the users own (similar to $P_i$ used in the labels of paper references). For each conflicting paper $i$, we use the disjunction $R_i \vee \texttt{\#CONFLICT}$ in the channel secrecy component to guarantee that a computation tainted with $R_i$ cannot write to the channel. Suppose a computation running on behalf of a reviewer in conflict with the $i$th paper reads review $R_i$. In this situation, the current label is set to $L = \langle R_i \wedge \cdots, \cdots \rangle$. Subsequent attempts to write to the output channel will be disallowed since $L \not\sqsubseteq L_o$. For $L \sqsubseteq L_o$ to hold, there must be a clause in $L_o$ that implies $R_i$. However, when in conflict, the only clause in the secrecy component that contains $R_i$ in the $R_i \vee \texttt{\#CONFLICT}$ (and clearly $R_i \vee \texttt{\#CONFLICT} \implies R_i$ does not hold).

[2] We loosely use the term "raise" to mean moving up the security lattice – this implies more secret, and of lower integrity in the DC label lattice.

### *3.3 Implementation*

Having established the underlying data structures and labeling patterns, we present the λChair API. As the main goal of λChair is to demonstrate the flexibility and power of our dynamic information flow library, we do not extend our example to a full-fledged system; the API can, however, be used to build relatively complex review systems. Below, we present the details of the λChair functions, which return actions in the `RevLIO` monad. As previously noted, this monad is a state monad with `LIO` as the base monad, threading the system users, review entries, and name of the current user through the computation.

**System administrator interface** A λChair administrator is provided with several functions that dynamically change the system state. From these functions, we detail the most interesting cases below.

- ▶ `addPaper :: Paper →RevLIO Id`
  Given a paper, it creates a new review entry for the paper and return the paper id. Internally, `addPaper` uses a function similar to `addReview` of Section 2.
- ▶ `addUser :: Name →Password →RevLIO ()`
  Given a unique user name and password, it adds the new user.
- ▶ `addAssignment :: Name →Id →RevLIO ()`
  Given a user name and paper id, it assigns the user to review the corresponding paper. The user must not be already in conflict with the paper.
- ▶ `addConflict :: Name →Id →RevLIO ()`
  Given a user name and paper id, it marks the user as being in conflict with the paper. As above, it must be the case that the user is not already assigned to review the paper.
- ▶ `asUser :: Name →RevLIO () →RevLIO ()`
  Given a user name, and user-constructed piece of code, it first authenticates the user and then executes the provided code with the current label and clearance of the user as described in Section 3.2. After the code is executed, the current label and clearance are restored and any information flow violations are reported.

**Reviewer interface** The reviewer, or user, composes an untrusted `RevLIO` computation (or action) that the trusted code executes using `asUser`. Such actions may be composed using the following interface:

- ▶ `findPaper :: String →RevLIO Id`
  Given a paper title, it returns its paper id, or fails if the paper is not found.
- ▶ `readPaper :: Id →RevLIO Paper`
  Given a paper id, the function returns an action which, when executed, returns the paper content.
- ▶ `readReview :: Id →RevLIO ()`
  Given a paper id, the function returns an action which, when executed, prints the review to the standard output. Its implementation is similar to the example of Section 2, except that it operates on references.
- ▶ `appendToReview :: Id→Content→RevLIO ()`
  Given a paper id and a review content, the function returns an action which, when executed, appends the supplied content to the review entry. Since there is no direct observation of the current review content, and to avoid label creep, the function, internally, uses `toLabeled`.

Figure 1 shows a simple example using the λChair API. In this example, Alice is assigned to review two papers. She does so by reading each paper (for the second, she also reads the existing reviews) and appending to the review "notebook". Bob, on the other hand, is added to the system after Alice's code is

```
module Admin where

import Alice
import Bob

main = evalRevLIO $ do
  -- Adding users to system
  addUser "Alice" "password"
  -- Adding papers to system
  p1 ← addPaper "Flexible Dynamic..."
  p2 ← addPaper "A Static..."
  -- Assign reviewers
  addAssignment "Alice" p1
  addAssignment "Alice" p2

  -- Executing Alice's code
  asUser "Alice" $ aliceCode

  -- Adding new users to system
  addUser "Bob" "password"
  -- Assign reviewers and conflicts
  addAssignment "Bob" p2
  addConflict "Bob"   p1

  -- Executing Bob's code
  asUser "Bob" $ bobCode
```

```
module Alice where

aliceCode = do
  p1 ← findPaper "Flexible Dynamic..."
  p2 ← findPaper "A Static..."
  readPaper p1
  appendToReview p1 "Interesting work!"
  readPaper p2
  readReview p2
  appendToReview p2 "What about adding
                     new users?"
  return ()
```

```
module Bob where

bobCode = do
  p1 ← findPaper "Flexible Dynamic..."
  p2 ← findPaper "A Static..."
  appendToReview p2 "Hmm, IFC.."
  readReview p1 -- IFC violation attempt
                -- (exception raised)
  return ()
```

Fig. 1: An example of code using λChair API.

executed. Bob first writes a review for the second paper and then attempts to violate IFC by trying to read (and write to the output channel) the reviews of the first paper. Though his review is appended to the correct paper, reading the review of the first paper is suppressed. Of course, the IFC violation attempt results in an exception. Though in this case Bob does not catch the exception, and the exception is propagated to the trusted API call `asUser` which handles the exceptions. Note, however, that Bob can safely recover from such IFC violation attempts. More specifically, the line `readReview p1` can be replaced by:

```
catch (readReview p1)
      (λ_ → writeToBobsLog "In conflict!" )
```

In this case, Bob's computation will terminate gracefully and simply write to log when he attempts to read the first paper's review.

## 4 Formal Semantics for LIO

This section formalizes our library for a call-by-name $\lambda$-calculus extended with Booleans, unit values, pairs, recursion, references, exceptions, and the `LIO` monadic operations. Figure 1 provides the formal syntax of the considered language. Syntactic categories $v$, $e$, and $\tau$ represent values, expressions, and types, respectively. Values are side-effect free while expressions denote (possible) side-effecting computations.

**Figure 1** Formal syntax for values, expressions, and types.

$$
\begin{array}{rrl}
\text{Value:} & v ::= & \texttt{true} \mid \texttt{false} \mid \texttt{()} \mid l \mid a \mid X \mid x \mid \lambda x.e \mid (e,e) \\
& & \mid \texttt{fix}\, e \mid \texttt{Lb}\, v\, e \mid (e)^{\text{LIO}} \mid X_l \mid \bullet \\
\text{Expression:} & e ::= & v \mid e\, e \mid \pi_i\, e \mid \texttt{if}\, e\, \texttt{then}\, e\, \texttt{else}\, e \\
& & \mid \texttt{let}\, x = e\, \texttt{in}\, e \mid \texttt{return}\, e \mid e \texttt{ >>= } e \\
& & \mid \texttt{label}\, e\, e \mid \texttt{unlabel}\, e \mid \texttt{toLabeled}\, e\, e \\
& & \mid \texttt{newLIORef}\, e\, e \mid \texttt{readLIORef}\, e \mid \texttt{writeLIORef}\, e\, e \\
& & \mid \texttt{throwLIO}\, e \mid \texttt{catch}\, e\, e \\
& & \mid \texttt{lowerClr}\, e \mid \texttt{getLabel} \mid \texttt{getClearance} \\
& & \mid \texttt{labelOf}\, e \mid \texttt{labelOfRef}\, e \\
\text{Type:} & \tau ::= & \texttt{Bool} \mid \texttt{()} \mid \tau \to \tau \mid (\tau,\tau) \mid \ell \\
& & \mid \texttt{Labeled}\, \ell\, \tau \mid \texttt{LIO}\, \ell\, \tau \mid \texttt{Ref}\, \ell\, \tau \mid \mathbf{X} \\
\text{Store:} & \phi : & \text{Address} \to \texttt{Labeled}\, \ell\, \tau
\end{array}
$$

**Values** In the syntax category $v$, symbol `true` and `false` represent Boolean values. Symbol `()` represents the unit value. Symbol $l$ denotes security labels. Symbol $a$ represents memory addresses in a given store. Symbol $X$ represents exceptions. Values include variables ($x$), functions ($\lambda x.e$), tuples $(e,e)$, and recursive functions ($\texttt{fix}\ e$). Four special syntax nodes are added to this category: $\texttt{Lb}\ v\ e$, $(e)^{\text{LIO}}$, $X_l$, and $\bullet$. Node $\texttt{Lb}\ v\ e$ denotes the run-time representation of a labeled value. Node $(e)^{\text{LIO}}$ denotes the run-time representation of a monadic `LIO` computation. Similarly node $X_l$ denotes the run-time representation of a labeled exception. Node $\bullet$ represents an erased value (explained in Section 5). We note that none of these special nodes appear in programs written by users and they are merely introduced for technical reasons.

**Expressions** Expressions are composed of values ($v$), function applications ($e\ e$), pair projections ($\pi_i\ e$), conditional branches ($\texttt{if}\ e\ \texttt{then}\ e\ \texttt{else}\ e$), and local definitions ($\texttt{let}\ x = e\ \texttt{in}\ e$). Additionally, expressions may involve operations related to monadic computations in the `LIO` monad. More precisely, $\texttt{return}\ e$ and $e \texttt{ >>= } e$ represent the monadic return and bind operations. Monadic operations related to the manipulation of labeled values inside the `LIO` monad are given by `label`, `unlabel`, and `toLabeled`. Expression $\texttt{label}\ e_1\ e_2$ creates a labeled value that guards $e_2$ with label $e_1$. Expression $\texttt{unlabel}\ e$ acquires the content of the labeled value $e$ while in a `LIO` computation. Expression $\texttt{toLabeled}\ e_1\ e_2$ creates a labeled value, with label $e_1$, of the result obtained by evaluating the `LIO` computation $e_2$. Non-proper morphisms related to creating, reading, and writing of references are respectively captured by expressions `newLIORef`, `readLIORef`, and `writeLIORef`. LIO operations may raise exceptions by calling `throwLIO` and catch exceptions with `catch`. Expression $\texttt{lowerClr}\ e$ allows lowering of the current clearance to $e$. Expressions `getLabel` and `getClearance` return the current label and current clearance of an `LIO` computation, respectively. Finally, expressions $\texttt{labelOf}\ e$ and $\texttt{labelOfRef}\ e$ respectively obtain the security label of labeled values and references.

**Types** We consider standard types for Booleans (`Bool`), unit (`()`), pairs $(\tau,\tau)$, and function $(\tau \to \tau)$ values. Type $\ell$ describes security labels. Type $\texttt{Labeled}\ \ell\ \tau$ describes labeled values of type $\tau$, where the

**Figure 2** Operational semantics for LIO (part I).

$$E ::= \cdots \mid \texttt{return}\, E \mid E \; \texttt{>>=} \; e \mid \texttt{labelOf}\, E \mid \texttt{labelOfRef}\, E$$

(RETURN)
$$\langle \Sigma, E[\texttt{return}\; v] \rangle \longrightarrow \langle \Sigma, E[(v)^{\text{LIO}}] \rangle$$

(BIND-1)
$$\langle \Sigma, E[(X_l)^{\text{LIO}} \; \texttt{>>=} \; e] \rangle \longrightarrow \langle \Sigma, E[(X_l)^{\text{LIO}}] \rangle$$

(BIND-2)
$$\frac{v \neq X_l}{\langle \Sigma, E[(v)^{\text{LIO}} \; \texttt{>>=} \; e] \rangle \longrightarrow \langle \Sigma, E[e\; v] \rangle}$$

(CLAB)
$$\frac{l = \Sigma.\texttt{lbl}}{\langle \Sigma, E[\texttt{getLabel}] \rangle \longrightarrow \langle \Sigma, E[\texttt{return}\; l] \rangle}$$

(CCLR)
$$\frac{l = \Sigma.\texttt{clr}}{\langle \Sigma, E[\texttt{getClearance}] \rangle \longrightarrow \langle \Sigma, E[\texttt{return}\; l] \rangle}$$

(GLAB)
$$\langle \Sigma, E[\texttt{labelOf}\; (\texttt{Lb}\; l\; e)] \rangle \longrightarrow \langle \Sigma, E[l] \rangle$$

(GLABR)
$$\frac{e = \Sigma.\phi(a)}{\langle \Sigma, E[\texttt{labelOfRef}\; a] \rangle \longrightarrow \langle \Sigma, E[\texttt{labelOf}\; e] \rangle}$$

label is of type $\ell$. Type `LIO` $\ell\ \tau$ represents monadic `LIO` computations, with a result type $\tau$ and the security labels of type $\ell$. Type `Ref` $\ell\ \tau$ describes labeled references, with labels of type $\ell$, to values of type $\tau$. Type **X** describes unlabeled exceptions[3].

### *4.1 Dynamic semantics for LIO*

The `LIO` monad presented in Section 2 is implemented as a state monad. Without loss of generality, we simplify the formalization and description of expressions by making the state of the monad part of a run-time environment. More precisely, for a given `LIO` computation, the symbol $\Sigma$ denotes a run-time environment that contains the current label, written $\Sigma.\texttt{lbl}$, the current clearance, written $\Sigma.\texttt{clr}$, and store, written $\Sigma.\phi$. We represent the store as a mapping from memory addresses ($a$) into labeled values ($\texttt{Lb}\; l\; e$). A run-time environment $\Sigma$ and `LIO` computation form a *configuration* $\langle \Sigma, e \rangle$. Given a configuration $\langle \Sigma, e \rangle$, the current label, clearance, and store when starting evaluation $e$ is given by $\Sigma.\texttt{lbl}$, $\Sigma.\texttt{clr}$, and $\Sigma.\phi$, respectively.

The relation $\langle \Sigma, e \rangle \longrightarrow \langle \Sigma', e' \rangle$ represents a single evaluation step from expression $e$, under the run-time environment $\Sigma$, to expression $e'$ and run-time environment $\Sigma'$—we say that $e$ reduces to $e'$ in one step. We write $\longrightarrow^*$ for the reflexive and transitive closure of $\longrightarrow$. The evaluation relation is defined in terms of a structured operational semantics via evaluation contexts (Felleisen, 1988).

The reduction rules for standard $\lambda$-calculus are self-explanatory and presented in Appendix A. More interestingly, Figures 2 and 3 present the non-standard evaluation contexts and reduction rules for our language. These rules guarantee that programs written using our approach fulfill non-interference, i.e., confidential information is not leaked, and confinement, i.e., a computation cannot access data above its clearance.

[3] For simplicity, we assume the set of exceptions is limited to a single type.

**Figure 3** Operational semantics for LIO (part II).

$$E ::= \cdots \mid \texttt{label}\, E\ e \mid \texttt{unlabel}\, E \mid \texttt{toLabeled}\, E\ e \mid \texttt{newLIORef}\, E\ e \mid \texttt{readLIORef}\, E \mid \texttt{writeLIORef}\, E\ e \mid \texttt{throwLIO}\, E \mid \texttt{catch}\, E\ e \mid \texttt{lowerClr}\, E$$

(LAB)

$$\frac{\Sigma.\texttt{lbl} \sqsubseteq l \sqsubseteq \Sigma.\texttt{clr}}{\langle \Sigma, E[\texttt{label}\ l\ e] \rangle \longrightarrow \langle \Sigma, E[\texttt{return}\ (\texttt{Lb}\ l\ e)] \rangle}$$

(UNLAB)

$$\frac{l' = \Sigma.\texttt{lbl} \sqcup l \qquad l' \sqsubseteq \Sigma.\texttt{clr} \qquad \Sigma' = \Sigma[\texttt{lbl} \mapsto l']}{\langle \Sigma, E[\texttt{unlabel}\ (\texttt{Lb}\ l\ e)] \rangle \longrightarrow \langle \Sigma', E[\texttt{return}\ e] \rangle}$$

(TOLAB-1)

$$\frac{\Sigma.\texttt{lbl} \sqsubseteq l \sqsubseteq \Sigma.\texttt{clr} \qquad \langle \Sigma, e \rangle \longrightarrow^* \langle \Sigma', (v)^{\text{LIO}} \rangle \qquad \Sigma'.\texttt{lbl} \sqsubseteq l \qquad \Sigma'' = \Sigma'[\texttt{lbl} \mapsto \Sigma.\texttt{lbl}, \texttt{clr} \mapsto \Sigma.\texttt{clr}]}{\langle \Sigma, E[\texttt{toLabeled}\ l\ e] \rangle \longrightarrow \langle \Sigma'', E[\texttt{label}\ l\ v] \rangle}$$

(TOLAB-2)

$$\frac{\Sigma.\texttt{lbl} \sqsubseteq l \sqsubseteq \Sigma.\texttt{clr} \qquad \langle \Sigma, e \rangle \longrightarrow^* \langle \Sigma', (v)^{\text{LIO}} \rangle \qquad \Sigma'.\texttt{lbl} \not\sqsubseteq l \qquad \Sigma'' = \Sigma'[\texttt{lbl} \mapsto \Sigma.\texttt{lbl}, \texttt{clr} \mapsto \Sigma.\texttt{clr}] \qquad l'' = l \sqcup \Sigma'.\texttt{lbl}}{\langle \Sigma, E[\texttt{toLabeled}\ l\ e] \rangle \longrightarrow \langle \Sigma'', E[\texttt{label}\ l\ X'_{l''}] \rangle}$$

(NREF)

$$\frac{\Sigma.\texttt{lbl} \sqsubseteq l \sqsubseteq \Sigma.\texttt{clr} \qquad \Sigma' = \Sigma.\phi[a \mapsto \texttt{Lb}\ l\ e]}{\langle \Sigma, E[\texttt{newLIORef}\ l\ e] \rangle \longrightarrow \langle \Sigma', E[\texttt{return}\ a] \rangle}\ a \textit{ fresh}$$

(RREF)

$$\frac{\Sigma.\phi(a) = \texttt{Lb}\ l\ e \qquad l' = \Sigma.\texttt{lbl} \sqcup l \qquad l' \sqsubseteq \Sigma.\texttt{clr} \qquad \Sigma' = \Sigma[\texttt{lbl} \mapsto l']}{\langle \Sigma, E[\texttt{readLIORef}\ a] \rangle \longrightarrow \langle \Sigma', E[\texttt{return}\ e] \rangle}$$

(WREF)

$$\frac{\Sigma.\phi(a) = \texttt{Lb}\ l\ e \qquad \Sigma.\texttt{lbl} \sqsubseteq l \sqsubseteq \Sigma.\texttt{clr} \qquad \Sigma' = \Sigma.\phi[a \mapsto \texttt{Lb}\ l\ e']}{\langle \Sigma, E[\texttt{writeLIORef}\ a\ e'] \rangle \longrightarrow \langle \Sigma', E[\texttt{return}\ ()] \rangle}$$

(THROW)

$$\frac{l = \Sigma.\texttt{lbl}}{\langle \Sigma, E[\texttt{throwLIO}\ X] \rangle \longrightarrow \langle \Sigma, E[(X_l)^{\text{LIO}}] \rangle}$$

(CATCH-1)

$$\frac{l' = \Sigma.\texttt{lbl} \sqcup l \qquad l' \sqsubseteq \Sigma.\texttt{clr} \qquad \Sigma' = \Sigma[\texttt{lbl} \mapsto l']}{\langle \Sigma, E[\texttt{catch}\ (X_l)^{\text{LIO}}\ e] \rangle \longrightarrow \langle \Sigma', E[e\ X] \rangle}$$

(CATCH-2)

$$\frac{v \neq X_l}{\langle \Sigma, E[\texttt{catch}\ (v)^{\text{LIO}}\ e] \rangle \longrightarrow \langle \Sigma, E[(v)^{\text{LIO}}] \rangle}$$

(LWCLR)

$$\frac{\Sigma.\texttt{lbl} \sqsubseteq l \sqsubseteq \Sigma.\texttt{clr} \qquad \Sigma' = \Sigma[\texttt{clr} \mapsto l]}{\langle \Sigma, E[\texttt{lowerClr}\ l] \rangle \longrightarrow \langle \Sigma', E[\texttt{return}\ ()] \rangle}$$

Rules in Figure 2 are self-explanatory, e.g., the evaluation rules for `return` and (`>>=`) are standard and labeled exceptions are propagated by (`>>=`) (rule (BIND-1)). Rule (LAB) of Figure 3 generates a labeled value if and only if the label is between the current label and clearance of the `LIO` computation. Rule (UNLAB) provides a method for accessing the content $e$ of a labeled value `Lb` $l$ $e$ in `LIO` computations. When the content of a labeled value is retrieved and used in an `LIO` computation, the current label is raised ($\Sigma' = \Sigma[\texttt{lbl} \mapsto l']$, where $l' = \Sigma.\texttt{lbl} \sqcup l$), capturing the fact that the remaining computation might depend on $e$. Of course, the current label should not exceed clearance ($l' \sqsubseteq \Sigma.\texttt{clr}$).

The reduction of `toLabeled` deserves some attention. Expression `toLabeled` $l$ $e$ is used to execute a computation $e$ to completion[4] ($\langle \Sigma, e \rangle \longrightarrow^* \langle \Sigma', (v)^{\text{LIO}} \rangle$) and wrap the result $v$ into a labeled value whose label is $l$. Specifying label $l$ is the responsibility of the programmer. We note, however, that the label $l$ needs to be an upper bound on the current label for the evaluation of computation $e$ ($\Sigma'.\texttt{lbl} \sqsubseteq l$), a restriction imposed in (TOLAB-1). The reason for this is due to the fact that security labels are protected by the current label, effectively making them public information accessible to any computation within scope (see rules (GLAB) and (GLABR)). As a consequence, a `toLabeled` that does not impose an upper bound on the sensitivity of the data observed by $e$ is susceptible to attacks. To illustrate this point, consider a computation with current label is $l_0$, that takes two (confidential) labeled values with respective labels $l_1$ and $l_2$ such that $l_i \not\sqsubseteq l_0, i = 1,2$. (Recall that the current label and clearance of a given `LIO` computation can be changed dynamically.) Further, suppose that `toLabeled` does not take an upper-bound on the computation's observations. Directly, the following program can be used to leak sensitive information:

```
leak lV1 lV2 = do          -- Initial label L_cur = l_0
  lV3 ← toLabeled $ do     -- Label of lV3 may be:
    v1 ← unlabel lV1       -- Read first value, raise label to L_cur = l_1
    if v1 then return True -- If value is, leave current label L_cur = l_1
          else unlabel lV2 -- Otherwise, the current label to L_cur = l_2
  return (labelOf lV3)     -- Can be l_1 or l_2
```

Note that, if the returned value of the inner computation can have the label $l_1$ or $l_2$ (remember that labels are effectively public information), information is directly leaked! Hence, to prevent such leaks, programmers must provide an upper-bound on the current label obtained when $e$ finishes computing. Since our approach is dynamic, flow-sensitive, and sound, this may require non-trivial static analysis in order to automatically determine the label for each call of `toLabeled` (Russo & Sabelfeld, 2010).

However, if the inner computation does read data more sensitive than $l$, such that the end current label $\Sigma'.\texttt{lbl} \not\sqsubseteq l$, rule (TOLAB-2) specifies that an exception labeled with the join of the upper-bound $l$ and $\Sigma'$ must be raised when performing an unlabel—hence, we return a labeled value that encloses an exception. Note that an exception is *not* raised at the point of evaluating `toLabeled`, but rather when the labeled value is unlabeled, and the current label is raised (see (UNLAB)).

When creating a reference, `newLIORef` $l$ $e$ produces a labeled value that guards $e$ with label $l$ and stores it in the memory store ($\Sigma' = \Sigma.\phi[a \mapsto \texttt{Lb}\ l\ e]$). The result of this operation is the memory address $a$ (`return` $a$). Observe that references are created only if the reference's label ($l$) is between the current label and clearance label ($\Sigma.\texttt{lbl} \sqsubseteq l \sqsubseteq \Sigma.\texttt{clr}$). As in (LAB), the restriction $l \sqsubseteq \Sigma.\texttt{clr}$ assures that programs cannot manipulate or access data beyond their clearance. Section 5 further details such confinement guarantees. Rule (RREF) obtains the content $e$ of a labeled value `Lb` $l$ $e$ stored in at address $a$. This rule raises the current label to the security level $l'$ ($\Sigma' = \Sigma[\texttt{lbl} \mapsto l']$ where $l' = \Sigma.\texttt{lbl} \sqcup l$). As in the previous rule, (RREF) enforces that the result of reading a reference is below the clearance ($l' \sqsubseteq \Sigma.\texttt{clr}$). Finally, rule (WREF) updates the memory store with a new value for the reference ($\Sigma' = \Sigma.\phi[a \mapsto \texttt{Lb}\ l\ e']$) as long as the label of the reference is above the current label and it does not exceed the clearance ($\Sigma.\texttt{lbl} \sqsubseteq l \sqsubseteq \Sigma.\texttt{clr}$).

[4] By using big-step semantics instead of an evaluation context of the form `toLabeled` $l$ $E$, the rules do not need to rely on the use of trusted primitives or a stack for (saving and) restoring the current label and clearance when executing `toLabeled`.

If considering $\Sigma.\texttt{lbl}$ as a dynamic version of the `pc` the restriction that the label of the reference must be above the current label ($\Sigma.\texttt{lbl} \sqsubseteq l$) is similar to the one imposed by (Pottier & Simonet, 2002).

Throwing and catching exceptions is standard. An `LIO` computation may raise an exception according to rule (THROW). The label of the raised exception is set to the current label. To handle exceptions raised in a computation $e_1$, a computation can execute the computation as $\texttt{catch}\ e_1\ e_2$, where $e_2$ corresponds to the exception handler. If no exception is raised, then rule (CATCH-2) simply propagates the value. However, if an exception is raised, and according to rule (CATCH-1), the current label is raised (clearance permitting) to the label of the exception and the exception handler is applied to the unlabeled exception. It is important to note that although our formalization of exceptions is limited to a single type, exceptions in LIO can encode information, similar to our encoding of the current label at the point of the throw.

Rule (LWCLR) allows a computation to lower the current clearance to $l$. This operation is particularly useful when trying to contain the access to some data as well as the effects produced by computations executed by `toLabeled`. Rules (CLAB) and (CCLR) obtain the current label and clearance from the run-time environment. Finally, rules (GLAB) and (GLABR) return the labels of labeled values and references. Observe that, regardless of the current label and clearance of the run-time environment, these two rules always succeed—hence “labels are public”.

**Addressing IFC violation attempts** Most of the evaluation rules in Figure 3 have a premise that imposes an information flow restriction. For example, rule (LAB) imposes the restriction that no labeled values may be labeled with a label below the current label or above the current clearance. As previously mentioned, rather than imposing that the evaluation of a misbehaving program gets “stuck”, we allow untrusted code to recover by throwing a monitor exception. Specifically, we introduce a “violation rule” for each rule that consists on the rule’s premise being negated and always evaluating to a `throwLIO`. For example, the violation rule for rule (LAB) is given by:

$$(\neg\text{LAB})\quad \frac{\neg(\Sigma.\texttt{lbl} \sqsubseteq l \sqsubseteq \Sigma.\texttt{clr})}{\langle \Sigma, E[\texttt{label}\ l\ e] \rangle \longrightarrow \langle \Sigma, E[\texttt{throwLIO}\ X] \rangle}$$

The remaining rules are similar and omitted for brevity.

### *4.2 Static semantics for LIO*

Figure 4 shows the typing rules for a subset of the terms and expressions; the remaining rules are shown in Appendix A. The typing rules are standard and we therefore do not describe them further. We note, however, that, unlike previous work (Russo *et al.*, 2008; Devriese & Piessens, 2011), we do not require the use of any sophisticated features from Haskell’s type-system, a direct consequence of our dynamic approach.

## 5 Soundness

In this section we show that `LIO` computations satisfy two security policies: non-interference and confinement. Non-interference shows that secrets are not leaked, while confinement establishes that certain pieces of code cannot manipulate or have access to certain data. The latter policy is similar to the confinement policies presented in (Leroy & Rouaix, 1998; Banerjee & Naumann, 2005).

**Figure 4** Typing rules for subset of terms and expressions.

$$\vdash l : \ell \qquad \frac{\Gamma(a) = \texttt{Labeled}\ \ell\ \tau}{\Gamma \vdash a : \texttt{Ref}\ \ell\ \tau} \qquad \Gamma \vdash X : \mathbf{X} \qquad \frac{\Gamma \vdash e_1 : \ell \qquad \Gamma \vdash e_2 : \tau}{\Gamma \vdash \texttt{Lb}\ e_1\ e_2 : \texttt{Labeled}\ \ell\ \tau} \qquad \frac{\Gamma \vdash e : \tau}{\Gamma \vdash (e)^{\texttt{LIO}} : \texttt{LIO}\ \ell\ \tau}$$

$$\frac{\Gamma \vdash e : \ell \qquad \Gamma \vdash X : \mathbf{X}}{\Gamma \vdash X_e : \tau} \qquad \Gamma \vdash \bullet : \tau \qquad \frac{\Gamma \vdash e_1 : \ell \qquad \Gamma \vdash e_2 : \tau}{\Gamma \vdash \texttt{label}\ e_1\ e_2 : \texttt{LIO}\ \ell\ (\texttt{Labeled}\ \ell\ \tau)}$$

$$\frac{\Gamma \vdash e : \texttt{Labeled}\ \ell\ \tau}{\Gamma \vdash \texttt{unlabel}\ e : \texttt{LIO}\ \ell\ \tau} \qquad \frac{\Gamma \vdash e_1 : \ell \qquad \Gamma \vdash e_2 : \texttt{LIO}\ \ell\ \tau}{\Gamma \vdash \texttt{toLabeled}\ e_1\ e_2 : \texttt{LIO}\ \ell\ (\texttt{Labeled}\ \ell\ \tau)}$$

$$\frac{\Gamma \vdash e_1 : \ell \qquad \Gamma \vdash e_2 : \tau}{\Gamma \vdash \texttt{newLIORef}\ e_1\ e_2 : \texttt{LIO}\ \ell\ (\texttt{Ref}\ \ell\ \tau)} \qquad \frac{\Gamma \vdash e : \texttt{Ref}\ \ell\ \tau}{\Gamma \vdash \texttt{readLIORef}\ e : \texttt{LIO}\ \ell\ \tau}$$

$$\frac{\Gamma \vdash e_1 : \texttt{Ref}\ \ell\ \tau \qquad \Gamma \vdash e_2 : \tau}{\Gamma \vdash \texttt{writeLIORef}\ e_1\ e_2 : \texttt{LIO}\ \ell\ ()} \qquad \frac{\Gamma \vdash e : \mathbf{X}}{\Gamma \vdash \texttt{throwLIO}\ e : \texttt{LIO}\ \ell\ \tau}$$

$$\frac{\Gamma \vdash e_1 : \texttt{LIO}\ \ell\ \tau \qquad \Gamma \vdash e_2 : \mathbf{X} \to \texttt{LIO}\ \ell\ \tau}{\Gamma \vdash \texttt{catch}\ e_1\ e_2 : \texttt{LIO}\ \ell\ \tau} \qquad \frac{\Gamma \vdash e : \ell}{\Gamma \vdash \texttt{lowerClr}\ e : \texttt{LIO}\ \ell\ ()} \qquad \vdash \texttt{getLabel} : \texttt{LIO}\ \ell\ \ell$$

$$\vdash \texttt{getClearance} : \texttt{LIO}\ \ell\ \ell \qquad \frac{\Gamma \vdash e : \texttt{Lb}\ \ell\ \tau}{\Gamma \vdash \texttt{labelOf}\ e : \ell} \qquad \frac{\Gamma \vdash e : \texttt{Ref}\ \ell\ \tau}{\Gamma \vdash \texttt{labelOfRef}\ e : \ell}$$

### *5.1 Non-interference*

As in (Li & Zdancewic, 2010; Russo *et al.*, 2008), we prove the non-interference property by using the technique of *term erasure*. Intuitively, data at security levels where the attacker cannot observe information can be safely rewritten to the syntax node $\bullet$. For the rest of the paper, we assume that the attacker can observe data up to security level $L$. The syntactic term $\bullet$, denoting an erased expression, may be associated to any type (recall Figure 9). Function $\varepsilon_L$ is responsible for performing the rewriting for data at security level not lower than $L$. In most of the cases, the erasure function is simply applied homomorphically (e.g., $\varepsilon_L(\texttt{if}\ E\ \texttt{then}\ e\ \texttt{else}\ e') = \texttt{if}\ \varepsilon_L(E)\ \texttt{then}\ \varepsilon_L(e)\ \texttt{else}\ \varepsilon_L(e')$). In the case of data constructors, it is simply the identity function. The definition of $\varepsilon_L$ for expressions and evaluation contexts are shown in Appendix B. Figure 5 shows the definition of $\varepsilon_L$ for terms, configurations, and bind. The three interesting cases for this function are when $\varepsilon_L$ is applied to a labeled value, a given configuration, or bind. In such cases, term erasing could indeed modify the behavior of the program. A labeled value is erased if the label assigned to it is above[5] $L$ ($\varepsilon_L(\texttt{Lb}\ l\ e) = \texttt{Lb}\ l\ \bullet$, if $l \not\sqsubseteq L$). Similarly, the computation performed in a certain configuration is erased if the current label is above $L$ ($\varepsilon_L(\langle \Sigma, e \rangle) = \langle \varepsilon_L(\Sigma), \bullet \rangle$ if $\Sigma.\texttt{lbl} \not\sqsubseteq L$). Finally, if $\varepsilon_L$ is applied to a bind-expression where the action evaluates to a labeled exception with label $l$ and $l \not\sqsubseteq L$, then the expression is fully erase to $(\bullet)^{\texttt{LIO}}$.

[5] We loosely use the word "above" to mean $\not\sqsubseteq$, since labels may not be comparable.

**Figure 5** Erasure function for terms, memory store, configurations and bind-expression.

$$\varepsilon_L(\texttt{true}) = \texttt{true} \qquad \varepsilon_L(\texttt{false}) = \texttt{false} \qquad \varepsilon_L(\texttt{()}) = \texttt{()} \qquad \varepsilon_L(l) = l \qquad \varepsilon_L(a) = a \qquad \varepsilon_L(x) = x$$

$$\varepsilon_L(\lambda x.e) = \lambda x.\varepsilon_L(e) \qquad \varepsilon_L((e,e)) = (\varepsilon_L(e), \varepsilon_L(e)) \qquad \varepsilon_L(\texttt{fix}\ e) = \texttt{fix}\ \varepsilon_L(e)$$

$$\varepsilon_L(\texttt{Lb}\ l\ e) = \begin{cases} \texttt{Lb}\ l\ \bullet & l \not\sqsubseteq L \\ \texttt{Lb}\ l\ \varepsilon_L(e) & \text{otherwise} \end{cases} \qquad \varepsilon_L((e)^{\texttt{LIO}}) = (\varepsilon_L(e))^{\texttt{LIO}} \qquad \varepsilon_L(\bullet) = \bullet$$

$$\frac{\varepsilon_L(\Sigma.\phi) = \{(x, \varepsilon_L(\Sigma.\phi(x)) : x \in \text{dom}(\Sigma.\phi)\}}{\varepsilon_L(\Sigma) = \Sigma[\phi \mapsto \varepsilon_L(\Sigma.\phi)]} \qquad \varepsilon_L(\langle \Sigma, e \rangle) = \begin{cases} \langle \varepsilon_L(\Sigma), \bullet \rangle & \Sigma.\texttt{lbl} \not\sqsubseteq L \\ \langle \varepsilon_L(\Sigma), \varepsilon_L(e) \rangle & \text{otherwise} \end{cases}$$

$$\varepsilon_L(X) = X \qquad \varepsilon_L(X_l) = \begin{cases} \bullet & l \not\sqsubseteq L \\ X_l & \text{otherwise} \end{cases}$$

$$\varepsilon_L(e_1 \texttt{ >>= } e_2) = \begin{cases} (\bullet)^{\texttt{LIO}} & e_1 = (X_l)^{\texttt{LIO}} \text{ and } l \not\sqsubseteq L \\ \varepsilon_L(e_1) \texttt{ >>= } \varepsilon_L(e_2) & \text{otherwise} \end{cases}$$

Following the definition of the erasure function, we introduce a new evaluation relation $\longrightarrow_L$ as follows:

*Definition 2* ($\longrightarrow_L$)

$$\frac{\langle \Sigma, e \rangle \longrightarrow \langle \Sigma', e' \rangle}{\langle \Sigma, e \rangle \longrightarrow_L \varepsilon_L(\langle \Sigma', e' \rangle)}$$

Expressions under this relationship are evaluated in the same way as before, with the exception that, after one evaluation step, the erasure function is applied to the resulting configuration, i.e., run-time environment and expression. In that manner, the relation $\longrightarrow_L$ guarantees that confidential data, i.e., data not below level $L$, is erased as soon as it is created. We write $\longrightarrow_L^*$ for the reflexive and transitive closure of $\longrightarrow_L$.

Most results that prove non-interference pursue the goal of establishing a relationship between $\longrightarrow^*$ and $\longrightarrow_L^*$ through the erasure function, as highlighted in Figure 6. Informally, the diagram establishes that erasing all secret data, i.e., data not below $L$, and then taking evaluation steps in $\longrightarrow_L$ is the same as taking steps in $\longrightarrow$ and then erasing all the secret values in the resulting configuration. Observe that if information from some level above $L$ is leaked by $e$, then erasing all secret data and then taking evaluation steps in $\longrightarrow_L$ might not be the same as taking steps in $\longrightarrow$ and then erasing all the secret values in the resulting configuration.

$$\begin{array}{ccc} \langle \Sigma, e \rangle & \longrightarrow^* & \langle \Sigma', e' \rangle \\ \downarrow \varepsilon_L & & \downarrow \varepsilon_L \\ \varepsilon_L(\langle \Sigma, e \rangle) & \longrightarrow_L^* & \varepsilon_L(\langle \Sigma', e' \rangle) \end{array}$$

Fig. 6: Simulation between $\longrightarrow^*$ and $\longrightarrow_L^*$.

For simplicity, we assume that the address space of the memory store is split into different security levels and that allocation is deterministic. In that manner, the address returned when creating a reference with level $l$ depends only on the references with level $l$ already in the store. These assumptions are valid in our language since, similar to traditional references in Haskell, we do not provide any mechanisms for deallocation or inspection of addresses in the API. However, when memory allocation is an observable channel, the library could be adapted in order to deal with non-opaque pointers (Hedin & Sands, 2006).

We start by showing that the evaluation relationships $\longrightarrow$ and $\longrightarrow_L$ are deterministic. Firstly, however, we note that $e = e'$ means syntactic equality between expressions $e$ and $e'$ and equality between run-time environments, written $\Sigma = \Sigma'$, is defined as the point-wise equality between mappings $\Sigma$ and $\Sigma'$.

*Proposition 1* (*Determinacy of* $\longrightarrow$)

- For any expression $e$ and run-time environment $\Sigma$ such that $\langle \Sigma, e \rangle \longrightarrow \langle \Sigma', e'' \rangle$, there is a unique term $e'$ and unique evaluation context $E$ such that $e = E[e']$.
- If $\langle \Sigma, e \rangle \longrightarrow \langle \Sigma', e' \rangle$ and $\langle \Sigma, e \rangle \longrightarrow \langle \Sigma'', e'' \rangle$, then $e' = e''$ and $\Sigma' = \Sigma''$.

*Proof*
By induction on expressions and evaluation contexts. □

*Proposition 2* (*Determinacy of* $\longrightarrow_L$)
If $\langle \Sigma, e \rangle \longrightarrow_L \langle \Sigma', e' \rangle$ and $\langle \Sigma, e \rangle \longrightarrow_L \langle \Sigma'', e'' \rangle$, then $e' = e''$ and $\Sigma' = \Sigma''$.

*Proof*
From Proposition 1 and definition of $\varepsilon_L$. □

The following proposition shows that the erasure function is homomorphic to the application of evaluation contexts and substitution as well as that it is idempotent.

*Proposition 3* (*Properties of erasure function*)

1. $\varepsilon_L(E[e]) = \varepsilon_L(E)[\varepsilon_L(e)]$
2. $\varepsilon_L([e_2/x]e_1) = [\varepsilon_L(e_2)/x]\varepsilon_L(e_1)$
3. $\varepsilon_L(\varepsilon_L(e)) = \varepsilon_L(e)$
4. $\varepsilon_L(\varepsilon_L(E)) = \varepsilon_L(E)$
5. $\varepsilon_L(\varepsilon_L(\Sigma)) = \varepsilon_L(\Sigma)$
6. $\varepsilon_L(\varepsilon_L(\langle \Sigma, e \rangle)) = \varepsilon_L(\langle \Sigma, e \rangle)$

*Proof*
Most cases follow by induction on expressions and evaluation contexts, see Appendix C for details. □

The next lemma establishes a simulation between $\longrightarrow$ and $\longrightarrow_L$ for expressions that do not execute `toLabeled`.

*Lemma 1* (*Single-step simulation without* `toLabeled`)
If $\Gamma \vdash e : \tau$ and $\langle \Sigma, e \rangle \longrightarrow \langle \Sigma', e' \rangle$ where `toLabeled` is not executed, then $\Gamma \vdash e' : \tau$ and $\varepsilon_L(\langle \Sigma, e \rangle) \longrightarrow_L \varepsilon_L(\langle \Sigma', e' \rangle)$.

*Proof*
Part of the lemma shows subject reduction, which is proved by showing that a reduction step does not change the types of references in the store $\Sigma.\phi$ and then applying induction on the typing derivations. The simulation follows by induction on evaluation contexts and case analysis on terms and expressions. Details are presented in Appendix C. □

Using this lemma, we then show that the simulation is preserved when performing several evaluation steps.

*Lemma 2* (*Simulation for expressions not executing* `toLabeled`)
If $\Gamma \vdash e : \tau$, $\langle \Sigma, e \rangle \longrightarrow^* \langle \Sigma', e' \rangle$ where there are no executions of `toLabeled`, then $\Gamma \vdash e' : \tau$ and $\varepsilon_L(\langle \Sigma, e \rangle) \longrightarrow_L^* \varepsilon_L(\langle \Sigma', e' \rangle)$.

*Proof*

By induction on $\longrightarrow$ and application of Lemma 1. $\square$

The reason for highlighting the distinction between expressions executing `toLabeled` and those not executing it is due to the fact that the evaluation of `toLabeled` involves big-step semantics (recall rules (TOLAB-1) and (TOLAB-2) in Figure 3). However, the next lemma shows the simulation between $\longrightarrow^*$ and $\longrightarrow_L^*$ for any expression $e$.

*Lemma 3* (*Simulation*)
If $\Gamma \vdash e : \tau$ and $\langle \Sigma, e \rangle \longrightarrow^* \langle \Sigma', e' \rangle$ then $\varepsilon_L(\langle \Sigma, e \rangle) \longrightarrow_L^* \varepsilon_L(\langle \Sigma', e' \rangle)$.

*Proof*
Lemma 2 shows the multi-step simulation for expressions that do not execute `toLabeled`. Thus, to show the general multi-step simulation, we first prove that `toLabeled` preserves the simulation by induction on the number of executed `toLabeled`. The general simulation follows directly. The interested reader is referred to Appendix C. $\square$

**Figure 7** *L*-equivalence for expressions.

$$\frac{e \approx_L e' \qquad l \sqsubseteq L}{\texttt{Lb}\ l\ e \approx_L \texttt{Lb}\ l\ e'} \qquad\qquad \frac{l \not\sqsubseteq L}{\texttt{Lb}\ l\ e \approx_L \texttt{Lb}\ l\ e'}$$

We define *L*-equivalence between expressions. Intuitively, two expressions are *L*-equivalent if they are syntactically equal, modulo labeled values whose labels do not flow to $L$. We use $\approx_L$ to represent *L*-equivalence for expressions. Figure 7 shows the definition for labeled values. Considering the simple lattice: $\texttt{L} \sqsubseteq \texttt{M} \sqsubseteq \texttt{H}$ and an attacker at level `L`, it holds that $\texttt{Lb H}\ 8 \approx_{\texttt{L}} \texttt{Lb H}\ 9$, but it does not hold that $\texttt{Lb L}\ 2 \approx_{\texttt{L}} \texttt{Lb L}\ 3$ or $\texttt{Lb H}\ 8 \approx_{\texttt{L}} \texttt{Lb M}\ 8$. Recall that labels are protected by the current label, and thus (usually) observable by an attacker — unlike the expressions they protect, labels must be the same even if they are above $L$. The rest of $\approx_L$ is defined as syntactic equality between constants (e.g., $\texttt{true} \approx_L \texttt{true}$) or homomorphisms (e.g., $\texttt{if}\ e\ \texttt{then}\ e_1\ \texttt{else}\ e_2 \approx_L \texttt{if}\ e'\ \texttt{then}\ e_1'\ \texttt{else}\ e_2'$ if $e \approx_L e'$, $e_1 \approx_L e_1'$, and $e_2 \approx_L e_2'$).

Since our language encompasses side-effecting expressions, it is also necessary to define *L*-equivalence between memory stores. Specifically, we say that two run-time environments are *L*-equivalent if an attacker at level $L$ cannot distinguish them:

*Definition 3* (*L-equivalence for stores*)

$$\frac{l \sqsubseteq L \vee l' \sqsubseteq L \qquad \forall a. \Sigma.\phi(a) = \texttt{Lb}\ l\ e \approx_L \Sigma'.\phi(a) = \texttt{Lb}\ l'\ e'}{\Sigma.\phi \approx_L \Sigma'.\phi}$$

Note that the *L*-equivalence ignores the store references with labels above $L$. Similarly, we define *L*-equivalence for configurations.

*Definition 4* (*L-equivalence for configurations*)

$$\frac{e \approx_L e' \qquad \Sigma.\phi \approx_L \Sigma'.\phi \qquad \Sigma.\texttt{lbl} = \Sigma'.\texttt{lbl} \qquad \Sigma.\texttt{clr} = \Sigma'.\texttt{clr} \qquad \Sigma.\texttt{lbl} \sqsubseteq L}{\langle \Sigma, e \rangle \approx_L \langle \Sigma', e' \rangle}$$

$$\frac{\Sigma.\phi \approx_L \Sigma'.\phi \qquad \Sigma.\texttt{lbl} \not\sqsubseteq L \qquad \Sigma'.\texttt{lbl} \not\sqsubseteq L}{\langle \Sigma, e \rangle \approx_L \langle \Sigma', e' \rangle}$$

In the above definition, it is worth remarking that we do not require $\approx_L$ for expressions when the current label does not flow to $L$. This omission comes from the fact that $e$ and $e'$ would be reduced to $\bullet$ when applying our simulation between $\longrightarrow^*$ and $\longrightarrow_L^*$ (recall Figure 5).

The next theorem shows the non-interference policy. It essentially states that given two inputs with possibly secret information, the result of the computation is indistinguishable to an attacker. In other words, there is no information-flow from confidential data to outputs observable by the attacker.

*Theorem 1* (*Non-interference*)
Given a computation $e$ (with no $\bullet$, $(\ )^{\text{LIO}}$, `Lb`, or $X_l$) where $\Gamma \vdash e : \texttt{Labeled}\ \ell\ \tau \rightarrow \texttt{LIO}\ \ell\ (\texttt{Labeled}\ \ell\ \tau')$, environments $\Sigma_1$ and $\Sigma_2$ where $\Sigma_1.\phi = \Sigma_2.\phi = \emptyset$, security label $l$, an attacker at level $L$ such that $l \sqsubseteq L$, then

$$\begin{aligned}
&\forall e_1 e_2.(\Gamma \vdash e_i : \texttt{Labeled}\ \ell\ \tau)_{i=1,2} \wedge (e_i = \texttt{Lb}\ l\ e_i')_{i=1,2} \wedge \langle \Sigma_1, e_1 \rangle \approx_L \langle \Sigma_2, e_2 \rangle \\
&\quad \wedge \langle \Sigma_1, e\ e_1 \rangle \longrightarrow^* \langle \Sigma_1', (v_1)^{\text{LIO}} \rangle \wedge \langle \Sigma_2, e\ e_2 \rangle \longrightarrow^* \langle \Sigma_2', (v_2)^{\text{LIO}} \rangle \\
&\quad \Longrightarrow \langle \Sigma_1', (v_1)^{\text{LIO}} \rangle \approx_L \langle \Sigma_2', (v_2)^{\text{LIO}} \rangle
\end{aligned}$$

*Proof*
From Lemma 3 and determinacy of $\longrightarrow_L^*$. The details are shown in Appendix C. □

Observe that even though we assume that the input labeled values $e_1$ and $e_2$ are observable by the attacker ($l \sqsubseteq L$), they might contain confidential data. For instance, $e_1$ could be of the form `Lb` $l$ (`Lb` $l'$ `true`) where $l' \not\sqsubseteq L$.

### *5.2 Confinement*

In this section we present the formal guarantees that `LIO` computations cannot modify data below their current label or above their current clearance.

We start by proving that the current label of an `LIO` computation does not decrease.

*Proposition 4* (*Monotonicity of the current label*)
If $\Gamma \vdash e : \tau$ and $\langle \Sigma, e \rangle \longrightarrow^* \langle \Sigma', e' \rangle$, then $\Sigma.\texttt{lbl} \sqsubseteq \Sigma'.\texttt{lbl}$.

*Proof*
By induction on expressions, evaluation contexts, and reduction rules. □

Similarly, we show that the current clearance of an `LIO` computation never increases.

*Proposition 5* (*Monotonicity of the current clearance*)
If $\Gamma \vdash e : \tau$ and $\langle \Sigma, e \rangle \longrightarrow^* \langle \Sigma', e' \rangle$, then $\Sigma'.\texttt{clr} \sqsubseteq \Sigma.\texttt{clr}$.

*Proof*
By induction on expressions, evaluation contexts, and reduction rules. □

Proposition 4 and 5 are crucial to assert that once an `LIO` computation reads confidential data, it cannot lower its current label to leak it. Similarly, a computation should not be able to arbitrarily increase its clearance; doing so would allow it to read any data with no access restrictions.

Before delving into the confinement theorems, we first define a store modifier that removes all store elements with a label that does not flow to $l$.

*Definition 5* (*Label-based reference-cell removal*)

Modifier $(\Sigma.\phi)_{\downarrow l}$ retains all the labeled references with a label below $l$, usually the current label:

$$(\Sigma.\phi)_{\downarrow l} = \Sigma.\phi \setminus \{(a, \texttt{Lb}\ l'\ e) : a \in \mathrm{dom}(\Sigma.\phi) \wedge l' \not\sqsubseteq l\}$$

And, dually, a store modifier that removes all store elements below a given clearance $l$.

*Definition 6* (*Clearance-based reference-cell removal*)

$$(\Sigma.\phi)_{\uparrow l} = \Sigma.\phi \setminus \{(a, \texttt{Lb}\ l'\ e) : a \in \mathrm{dom}(\Sigma.\phi) \wedge l \not\sqsubseteq l'\}$$

This store modifier retains all the labeled references with a label that is not below $l$, usually the current clearance. We now present the first confinement theorem.

*Theorem 2* (*Store confinement*)
Given labels $l$ and $l_c$, a computation $e$ (with no $\bullet$, $a$, $(\ )^{\text{LIO}}$, $\texttt{Lb}$, or $X_{l'}$) such that $\Gamma \vdash e : \texttt{LIO}\ \ell\ \tau$, and environment $\Sigma[\texttt{lbl} \mapsto l, \texttt{clr} \mapsto l_c]$ where $l \sqsubseteq l_c$, then

$$\langle \Sigma, e \rangle \longrightarrow^{*} \langle \Sigma', (v)^{\text{LIO}} \rangle \implies (\Sigma.\phi)_{\downarrow l} = (\Sigma'.\phi)_{\downarrow l} \wedge (\Sigma.\phi)_{\uparrow l_c} = (\Sigma'.\phi)_{\uparrow l_c}$$

*Proof*
By contradiction on creating and modifying labeled references with labels not boded by the current label and clearance, using Propositions 4 and 5. □

Intuitively, this theorem states that no new references with a label not bounded by the initial current label and current clearance can be created. And, computation $e$ is confined to modifying references between $l$ and $l_c$.

Our second confinement theorem states that a return labeled value comes either from some part of the store (recall that labeled values can be nested) or might be computed when its security level is between the current label and clearance.

*Theorem 3* (*$\texttt{Labeled}$ creation confinement*)
Given labels $l$, $l_c$, and $l_v$, a computation $e$ (with no $\bullet$, $a$, $(\ )^{\text{LIO}}$, $\texttt{Lb}$, or $X_{l'}$) where $\Gamma \vdash e : \texttt{LIO}\ \ell\ (\texttt{Labeled}\ \ell\ \tau)$, and environment $\Sigma[\texttt{lbl} \mapsto l, \texttt{clr} \mapsto l_c]$ such that $l \sqsubseteq l_c$, then

$$\langle \Sigma, e \rangle \longrightarrow^{*} \langle \Sigma', (\texttt{Lb}\ l_v\ e_1)^{\text{LIO}} \rangle \implies l \sqsubseteq l_v \sqsubseteq l_c \vee \exists (a, \texttt{Lb}\ l_1\ e_1') \in \Sigma.\phi . \texttt{Lb}\ l_v\ e_1\ \tilde{\in}\ e_1' \wedge l_1 \sqsubseteq l_c$$

Here, operator $\tilde{\in}$ is defined as the syntactic appearance of the left-hand expression into the right-hand side operand.

*Proof*
By induction on expressions and evaluation contexts and using Propositions 4 and 5. □

## 6 Related Work

Heintze and Riecke (Heintze & Riecke, 1998) consider security for lambda-calculus where lambda-terms are explicitly annotated with security labels, for a type-system that guarantees non-interference. One of the key aspects of their work consists of an operator which raises the security annotation of a term in a similar manner to our raise of the current label when manipulating labeled values. Similar ideas of floating labels have been used by many operating systems, dating back to the High-Water-Mark security model (Landwehr, 1981) of the ADEPT-50 in the late 1960s. Asbestos (Efstathopoulos *et al.*, 2005) first combined floating labels with the Decentralized label model (Myers & Liskov, 1997).

Abadi et al. (Abadi *et al.*, 1999) develop the dependency core calculus (DCC) based on a hierarchy of monads to guarantee non-interference. In their calculus, they define a monadic type that "protects" (the confidentiality of) side-effect-free values at different security levels. Though not a monad, our `Labeled` type similarly protects pure values at various security levels. To manipulate such values, DCC uses a non-standard typing rule for the bind operator; the essence of this operator, in a dynamic setting with side-effectful computations, is captured in our library through the interaction of of `Labeled`, `unlabel`, and `LIO`.

Tse and Zdancewic (Tse & Zdancewic, 2004) translate DCC to System F and show that non-interference can be stated using the parametricity theorem for System F. The authors also provide a Haskell implementation for a two-point lattice. Their implementation encodes each security level as an abstract data type constructed from functions and binding operations to compose computations with permitted flows. Since they consider the same non-standard features for the `bind` operation as in DCC, they provide as many definitions for `bind` as different type of values produced by it. Moreover, their implementation needs to be compiled with the flag `-fallow-undecidable-instances`, in GHC. Our work, in contrast, defines only one bind operation for `LIO`, without the need for such compiler extensions.

Harrison and Hook show how to implement an abstract operating system called *separation kernel* (Harrison, 2005). Programs running under this multi-threading operating system satisfy non-interference. To achieve this, the authors rely on the state monad to represent threads, monad transformers to present parallel composition, and the resumption monad to achieve communication between threads. Consequently, non-interference is enforced by the scheduler implementation, which only allow signaling threads at the same, or higher, security level as the thread that issued the signal. The authors use monads differently from us; their goal is to construct secure kernels rather than provide information-flow security as a library. Our library is simpler and more suitable for writing sequential programs in Haskell. Extending our library to include concurrency is stated as a future work.

Crary et al. (Crary *et al.*, 2005) design a monadic calculus for non-interference for programs with mutable state. Similar to our work, their language distinguishes between term and expressions, where terms are pure and expressions are (possibly) effectful computations. Their calculus mainly tracks information flow by statically approximating the security levels of effects produced by expressions. Compared to their work, we only need to make approximations of the side-effects of a given computation when using `toLabeled`; the state of `LIO` keeps track of the dynamic security level upper bound of observed data. Overall, our dynamic approach is more flexible and permissive than their proposed type-system.

Pottier and Simonet (Pottier & Simonet, 2002; Simonet, 2003) designed FlowCaml, a compiler to enforce non-interference for OCaml programs. Rather than implementing a compiler from scratch, and more similar to our approach, the seminal work by Li and Zdancewic (Li & Zdancewic, 2006) presents an implementation of information-flow security as a library, in Haskell, using a generalization of monads called Arrows (Hughes, 2000). Extending their work, Tsai et al. (chung Tsai *et al.*, 2007) further consider side-effects and concurrency. Contributing to library-based approaches, Russo et al. (Russo *et al.*, 2008) eliminate the need for Arrows by showing an IFC library based solely on monads. Their library defines monadic types to track information-flow in pure and side-effectful computations. Compared to our dynamic IFC library, Russo et al.'s library is slightly less permissive and leverages Haskell's type-system to statically enforce non-interference. However, we note that our library has similar (though dynamic) functions provided by their SecIO library; similar to `unlabel`, they provide a function that maps pure labeled values into side-effectful computations; similar to `toLabeled`, they provide a function that allows reading/writing secret files into computations related to public data.

Morgenstern et al. (Morgenstern & Licata, 2010) encoded an authorization- and IFC-aware programming language in Agda. Their encoding, however, does not consider computations with side-effects. More closely related, Devriese and Piessens (Devriese & Piessens, 2011) used monad transformers and parametrized monads (Atkey, 2006) to enforce non-interference, both dynamically and statically. However, their work focuses on modularity (separating IFC enforcement from underlying user API), using type-class level tricks that make it difficult to understand errors triggered by insecurities. Moreover, compared to our work, where programmers write standard Haskell code, their work requires one to firstly encode programs as values of a specific type.

Compared to other language-based works, LIO uses the notion of clearance. The work of Bell and La Padula (Bell & Padula, 1976) formalized clearance as a bound on the current label of a particular users' processes. In the 1980s, clearance became a requirement for high-assurance secure systems purchased by the US Department of Defense (Department of Defense, 1985). More recently, HiStar (Zeldovich *et al.*, 2006) re-cast clearance as a bound on the label of any resource created by the process (where raising a process's label is but one means of creating a something with a higher label). We adopt HiStar's more stringent notion of clearance, which prevents software from copying data it cannot read and facilitates bounding the time during which possibly untrustworthy software can exploit covert channels.

Simultaneously to this work, Hedin and Sabelfeld have recently published a dynamic information-flow monitor for a core part of Javascript which handles exceptions (Hedin & Sabelfeld, 2012). Their approach needs to explicitly mark in the code (through non-standard constructors) which exceptions are thrown under a secret branch. Our approach, in contrast, simple raises an exception labeled with the current label.

## 7 Conclusion

We propose a new design point for IFC systems in which most values in lexical scope are protected by a single, mutable, *current label*, yet one can also encapsulate and pass around the results of computations with different labels. Unlike other language-based work, our model provides a notion of *clearance* that imposes an upper bound on the program label, thus providing a form of discretionary access control on portions of the code.

We prove information flow and integrity properties of our design and describe LIO, an implementation of the new model in Haskell. LIO, which can be implemented entirely as a library, demonstrates both the applicability and simplicity of the approach. We show the capabilities of the library to perform secure side-effects (e.g., references) as well as safely handle exceptions. Our non-interference theorem proves the conventional property that lower-level results do not depend on higher-level inputs – the label system prevents inappropriate flow of information. We also prove confinement theorems that show the effect of clearance on the behavior of code. In effect, lowering the clearance imposes a discretionary form of access control by preventing subsequent code (within that scope) from accessing higher-level information.

As an illustration of the benefits and expressive power of this system, we describe a reviewing system that uses LIO labels to manage integrity and confidentiality in an environment where users and labels are added dynamically. Although we have use LIO for the $\lambda$Chair API and even built a relatively large web-framework that securely integrates untrusted third-party applications, we believe that changes in the constructs are likely to occur as the language matures. This further supports our library-based approach to language-based security.

**Acknowledgments** This work was funded by DARPA CRASH under contract #N66001-10-2-4088, by multiple gifts from Google, and by the Swedish research agencies VR and STINT. D. Stefan is supported by the DoD through the NDSEG Fellowship Program.

## A Standard Static and Dynamic Semantics

For completeness, in this section, we provide the evaluation and typing rules for standard terms and expressions. Figure 8 defines the set of evaluation contexts and reduction rules for standard constructs in our language. Substitution ($[e_1/x]\ e_2$) is defined in the usual way: homomorphic on all operators and renaming bound names to avoid captures. Figure 9 describes the typing rules for terms; Figure 10 describes the typing rules for expressions.

**Figure 8** Operational semantics for standard terms.

$$E ::= [\cdot] \mid \texttt{Lb}\ E\ e \mid E\ e \mid \pi_i\ E \mid \texttt{if}\ E\ \texttt{then}\ e\ \texttt{else}\ e$$

$$\begin{array}{l}
\langle \Sigma, E[(\lambda x.e_1)\ e_2]\rangle \longrightarrow \langle \Sigma, E[[e_2/x]e_1]\rangle \\
\langle \Sigma, E[\texttt{fix}\ e]\rangle \longrightarrow \langle \Sigma, E[e\ (\texttt{fix}\ e)]\rangle \\
\langle \Sigma, E[\pi_i\ (e_1, e_2)]\rangle \longrightarrow \langle \Sigma, E[e_i]\rangle \\
\langle \Sigma, E[\texttt{if}\ \texttt{true}\ \texttt{then}\ e_1\ \texttt{else}\ e_2]\rangle \longrightarrow \langle \Sigma, E[e_1]\rangle \\
\langle \Sigma, E[\texttt{if}\ \texttt{false}\ \texttt{then}\ e_1\ \texttt{else}\ e_2]\rangle \longrightarrow \langle \Sigma, E[e_2]\rangle \\
\langle \Sigma, E[\texttt{let}\ x = e_1\ \texttt{in}\ e_2]\rangle \longrightarrow \langle \Sigma, E[[e_1/x]e_2]\rangle
\end{array}$$

**Figure 9** Typing rules for standard terms.

$$\vdash \texttt{true} : \texttt{Bool} \qquad \vdash \texttt{false} : \texttt{Bool} \qquad \vdash () : () \qquad \frac{\Gamma(x) = \tau}{\Gamma \vdash x : \tau} \qquad \frac{\Gamma[x \mapsto \tau_1] \vdash e : \tau_2}{\Gamma \vdash \lambda x.e : \tau_1 \to \tau_2}$$

$$\frac{\Gamma \vdash e_1 : \tau_1 \qquad \Gamma \vdash e_2 : \tau_2}{\Gamma \vdash (e_1, e_2) : (\tau_1, \tau_2)} \qquad \frac{\Gamma \vdash e : \tau \to \tau}{\Gamma \vdash \texttt{fix}\ e : \tau} \qquad \frac{\Gamma \vdash e : \tau}{\Gamma \vdash (e)^{\text{LIO}} : \texttt{LIO}\ \ell\ \tau} \qquad \Gamma \vdash \bullet : \tau$$

**Figure 10** Typing rules for standard expressions.

$$\frac{\Gamma \vdash e_1 : \tau_1 \to \tau_2 \qquad \Gamma \vdash e_2 : \tau_1}{\Gamma \vdash e_1\ e_2 : \tau_2} \qquad \frac{\Gamma \vdash e : (\tau_1, \tau_2)}{\Gamma \vdash \pi_i\ e : \tau_i} \qquad \frac{\Gamma \vdash e_1 : \texttt{Bool} \qquad \Gamma \vdash e_2 : \tau \qquad \Gamma \vdash e_3 : \tau}{\Gamma \vdash \texttt{if}\ e_1\ \texttt{then}\ e_2\ \texttt{else}\ e_3 : \tau}$$

$$\frac{\Gamma \vdash e_1 : \tau_1 \qquad \Gamma[x \mapsto \tau_1] \vdash e_2 : \tau_2}{\Gamma \vdash \texttt{let}\ x = e_1\ \texttt{in}\ e_2 : \tau_2} \qquad \frac{\Gamma \vdash e : \tau}{\Gamma \vdash \texttt{return}\ e : \texttt{LIO}\ \ell\ \tau}$$

$$\frac{\Gamma \vdash e_1 : \texttt{LIO}\ \ell\ \tau_1 \qquad \Gamma \vdash e_2 : \tau_1 \to \texttt{LIO}\ \ell\ \tau_2}{\Gamma \vdash e_1\ \texttt{>>=}\ e_2 : \texttt{LIO}\ \ell\ \tau_2}$$

## B Erasure function

In this section we define the erasure function $\varepsilon_L$, introduced in Section 5 for the remaining expressions (Figure 11) and evaluation contexts (Figure 12).

**Figure 11** Erasure function for expressions.

$$\varepsilon_L(e\ e) = \varepsilon_L(e)\ \varepsilon_L(e) \qquad \varepsilon_L(\pi_i\ e) = \pi_i\ \varepsilon_L(e)$$

$$\varepsilon_L(\texttt{if}\ e\ \texttt{then}\ e\ \texttt{else}\ e) = \texttt{if}\ \varepsilon_L(e)\ \texttt{then}\ \varepsilon_L(e)\ \texttt{else}\ \varepsilon_L(e)$$

$$\varepsilon_L(\texttt{let}\ x = e\ \texttt{in}\ e) = \texttt{let}\ x = \varepsilon_L(e)\ \texttt{in}\ \varepsilon_L(e) \qquad \varepsilon_L(\texttt{return}\ e) = \texttt{return}\ \varepsilon_L(e)$$

$$\varepsilon_L(\texttt{label}\ e\ e) = \texttt{label}\ \varepsilon_L(e)\ \varepsilon_L(e) \qquad \varepsilon_L(\texttt{unlabel}\ e) = \texttt{unlabel}\ \varepsilon_L(e)$$

$$\varepsilon_L(\texttt{toLabeled}\ e\ e) = \texttt{toLabeled}\ \varepsilon_L(e)\ \varepsilon_L(e) \qquad \varepsilon_L(\texttt{newLIORef}\ e\ e) = \texttt{newLIORef}\ \varepsilon_L(e)\ \varepsilon_L(e)$$

$$\varepsilon_L(\texttt{readLIORef}\ e) = \texttt{readLIORef}\ \varepsilon_L(e) \qquad \varepsilon_L(\texttt{writeLIORef}\ e\ e) = \texttt{writeLIORef}\ \varepsilon_L(e)\ \varepsilon_L(e)$$

$$\varepsilon_L(\texttt{throwLIO}\ e) = \texttt{throwLIO}\ \varepsilon_L(e) \qquad \varepsilon_L(\texttt{catch}\ e\ e) = \texttt{catch}\ \varepsilon_L(e)\ \varepsilon_L(e)$$

$$\varepsilon_L(\texttt{lowerClr}\ e) = \texttt{lowerClr}\ \varepsilon_L(e) \qquad \varepsilon_L(\texttt{getLabel}) = \texttt{getLabel}$$

$$\varepsilon_L(\texttt{getClearance}) = \texttt{getClearance} \qquad \varepsilon_L(\texttt{labelOf}\ e) = \texttt{labelOf}\ \varepsilon_L(e)$$

$$\varepsilon_L(\texttt{labelOfRef}\ e) = \texttt{labelOfRef}\ \varepsilon_L(e)$$

**Figure 12** Erasure function for evaluation contexts.

$$\varepsilon_L(\texttt{Lb}\ E\ e) = \texttt{Lb}\ \varepsilon_L(E)\ \varepsilon_L(e) \qquad \varepsilon_L(E\ e) = \varepsilon_L(E)\ \varepsilon_L(e) \qquad \varepsilon_L(\pi_i\ E) = \pi_i\ \varepsilon_L(E)$$

$$\varepsilon_L(\texttt{if}\ E\ \texttt{then}\ e\ \texttt{else}\ e) = \texttt{if}\ \varepsilon_L(E)\ \texttt{then}\ \varepsilon_L(e)\ \texttt{else}\ \varepsilon_L(e) \qquad \varepsilon_L(\texttt{return}\ E) = \texttt{return}\ \varepsilon_L(E)$$

$$\varepsilon_L(E\ \texttt{>>=}\ e) = \varepsilon_L(E)\ \texttt{>>=}\ \varepsilon_L(e) \qquad \varepsilon_L(\texttt{label}\ E\ e) = \texttt{label}\ \varepsilon_L(E)\ \varepsilon_L(e)$$

$$\varepsilon_L(\texttt{unlabel}\ E) = \texttt{unlabel}\ \varepsilon_L(E) \qquad \varepsilon_L(\texttt{toLabeled}\ E\ e) = \texttt{toLabeled}\ \varepsilon_L(E)\ \varepsilon_L(e)$$

$$\varepsilon_L(\texttt{newLIORef}\ E\ e) = \texttt{newLIORef}\ \varepsilon_L(E)\ \varepsilon_L(e) \qquad \varepsilon_L(\texttt{readLIORef}\ E) = \texttt{readLIORef}\ \varepsilon_L(E)$$

$$\varepsilon_L(\texttt{writeLIORef}\ E\ e) = \texttt{writeLIORef}\ \varepsilon_L(E)\ \varepsilon_L(e) \qquad \varepsilon_L(\texttt{throwLIO}\ E) = \texttt{throwLIO}\ \varepsilon_L(E)$$

$$\varepsilon_L(\texttt{catch}\ E\ e) = \texttt{catch}\ \varepsilon_L(E)\ \varepsilon_L(e) \qquad \varepsilon_L(\texttt{lowerClr}\ E) = \texttt{lowerClr}\ \varepsilon_L(E)$$

$$\varepsilon_L(\texttt{labelOf}\ E) = \texttt{labelOf}\ \varepsilon_L(E) \qquad \varepsilon_L(\texttt{labelOfRef}\ E) = \texttt{labelOfRef}\ \varepsilon_L(E)$$

## C Detailed proofs

In this section, we provide expand the proof details for the results in Section 5.

*Proposition 3* (*Properties of erasure function*)

1. $\varepsilon_L(E[e]) = \varepsilon_L(E)[\varepsilon_L(e)]$
2. $\varepsilon_L([e_2/x]e_1) = [\varepsilon_L(e_2)/x]\varepsilon_L(e_1)$
3. $\varepsilon_L(\varepsilon_L(e)) = \varepsilon_L(e)$
4. $\varepsilon_L(\varepsilon_L(E)) = \varepsilon_L(E)$
5. $\varepsilon_L(\varepsilon_L(\Sigma)) = \varepsilon_L(\Sigma)$
6. $\varepsilon_L(\varepsilon_L(\langle\Sigma, e\rangle)) = \varepsilon_L(\langle\Sigma, e\rangle)$

*Proof*

All follow from the definition of the erasure function $\varepsilon_L$, and induction on expressions and evaluation contexts,

1. By induction on expressions and evaluation contexts. We show several cases of the base case analysis on evaluation contexts.
   (a) Let $E := \texttt{Lb}\ [\,]\ e_0$, it follows that $\varepsilon_L(E) := \texttt{Lb}\ [\,]\ \varepsilon_L(e_0)$, and $\varepsilon_L(E[e]) = \varepsilon_L(\texttt{Lb}\ e\ e_0) = \texttt{Lb}\ \varepsilon_L(e)\ \varepsilon_L(e_0) = \varepsilon_L(E)[\varepsilon_L(e)]$.
   (b) Let $E := [\,]\ e_0$, it follows that $\varepsilon_L(E) := [\,]\ \varepsilon_L(e_0)$, and $\varepsilon_L(E[e]) = \varepsilon_L(e\ e_0) = \varepsilon_L(e)\ \varepsilon_L(e_0) = \varepsilon_L(E)[\varepsilon_L(e)]$.
   (c) Let $E := \pi_i\ [\,]$, it follows that $\varepsilon_L(E) := \pi_i\ [\,]$, and $\varepsilon_L(E[e]) = \varepsilon_L(\pi_i\ e) = \pi_i\ \varepsilon_L(e) = \varepsilon_L(E)[\varepsilon_L(e)]$.
2. By expansion $\varepsilon_L([e_2/x]e_1) = \varepsilon_L((\lambda x.e_1)\ e_2)$, from which we have $\varepsilon_L(\lambda x.e_1)\ \varepsilon_L(e_2) = [\varepsilon_L(e_2)/x]\varepsilon_L(e_1)$.
3. Directly from definition of the erasure function and induction on expressions.
4. Directly from definition of the erasure function and induction on expressions and evaluation contexts.
5. Directly from definition of the erasure function on stores and property 3 above.
6. Directly from definition of the erasure function on configurations and properties 3 and 5, above.

□

*Lemma 1* (*Single-step simulation without* `toLabeled`)

If $\Gamma \vdash e : \tau$ and $\langle\Sigma, e\rangle \longrightarrow \langle\Sigma', e'\rangle$ where `toLabeled` is not executed, then $\Gamma \vdash e' : \tau$ and $\varepsilon_L(\langle\Sigma, e\rangle) \longrightarrow_L \varepsilon_L(\langle\Sigma', e'\rangle)$.

*Proof*

Part of the lemma shows subject reduction, which is proved by showing that a reduction step does not change the types of references in the store $\Sigma.\phi$ and then applying induction on the typing derivations.

It remains then to show the simulation, which follows by induction on evaluation contexts and cases analysis on terms and expressions. For clarity, we omit the environment in cases where it is not essential. Unless otherwise stated, we assume that $\Sigma.\texttt{lbl} \sqsubseteq L \sqsubseteq \Sigma.\texttt{clr}$, the proof for the case where $L$ is below the current label is straight forward since the $\varepsilon_L$ erases any expression in a configuration to a hole. Conversely, the case where $L$ is above the current clerance is identical to the case where $L$ is equal to the current clearance.

We show the simulation for several exemplary/interesting cases, the remaining cases follow similarly.

▶ Case $E[(\lambda x.e_1)\ e_2] \longrightarrow E[[e_2/x]e_1]$:

$$\begin{aligned}
& \varepsilon_L(E[(\lambda x.e_1)\ e_2]) = \varepsilon_L(E)[\varepsilon_L((\lambda x.e_1)\ e_2)] \\
& = \varepsilon_L(E)[(\lambda x.\varepsilon_L(e_1))\ \varepsilon_L(e_2)] \\
\longrightarrow_L\ & \varepsilon_L(\varepsilon_L(E)[[\varepsilon_L(e_2)/x]\varepsilon_L(e_1)]) \\
& = \varepsilon_L(\varepsilon_L(E))[\varepsilon_L([\varepsilon_L(e_2)/x]\varepsilon_L(e_1))] \\
& = \varepsilon_L(E)[\varepsilon_L([\varepsilon_L(e_2)/x]\varepsilon_L(e_1))] \\
& = \varepsilon_L(E)[[\varepsilon_L(e_2)/x]\varepsilon_L(e_1)] \\
& = \varepsilon_L(E)[\varepsilon_L([e_2/x]e_1)] = \varepsilon_L(E[[e_2/x]e_1])
\end{aligned}$$

by Proposition 3.

▶ Case $\langle \Sigma, E[\texttt{return}\ v] \rangle \longrightarrow \langle \Sigma, E[(v)^{\text{LIO}}] \rangle$:

— $\Sigma.\texttt{lbl} \sqsubseteq L$:

$$\begin{aligned}
& \varepsilon_L(\langle \Sigma, E[\texttt{return}\ v] \rangle) \\
& = \langle \varepsilon_L(\Sigma), \varepsilon_L(E[\texttt{return}\ v]) \rangle \\
& = \langle \varepsilon_L(\Sigma), \varepsilon_L(E)[\texttt{return}\ \varepsilon_L(v)] \rangle \\
\longrightarrow_L\ & \varepsilon_L(\langle \varepsilon_L(\Sigma), \varepsilon_L(E)[(\varepsilon_L(v))^{\text{LIO}}] \rangle) \\
& = \langle \varepsilon_L(\Sigma), \varepsilon_L(E)[(\varepsilon_L(v))^{\text{LIO}}] \rangle \\
& = \langle \varepsilon_L(\Sigma), \varepsilon_L(E)[\varepsilon_L((v)^{\text{LIO}})] \rangle \\
& = \langle \varepsilon_L(\Sigma), \varepsilon_L(E[(v)^{\text{LIO}}]) \rangle = \varepsilon_L(\langle \Sigma, E[(v)^{\text{LIO}}] \rangle)
\end{aligned}$$

by definition of $\varepsilon_L$ and Proposition 3.

— $\Sigma.\texttt{lbl} \not\sqsubseteq L$:

$$\begin{aligned}
& \varepsilon_L(\langle \Sigma, E[\texttt{return}\ v] \rangle) = \langle \varepsilon_L(\Sigma), \bullet \rangle \\
\longrightarrow_L\ & \varepsilon_L(\langle \varepsilon_L(\Sigma), \bullet \rangle) = \langle \varepsilon_L(\Sigma), \bullet \rangle = \varepsilon_L(\langle \Sigma, E[(v)^{\text{LIO}}] \rangle)
\end{aligned}$$

by definition of $\varepsilon_L$ and Proposition 3.

This illustrates the approach used to prove simulation of most cases. Moreover, it shows the trivial case for $\Sigma.\texttt{lbl} \not\sqsubseteq L$.

▶ Case $\langle \Sigma, E[(X_l)^{\text{LIO}} \texttt{>>=}\ e] \rangle \longrightarrow \langle \Sigma, E[(X_l)^{\text{LIO}}] \rangle$

— $l \sqsubseteq L$:

$$\begin{aligned}
& \varepsilon_L(\langle \Sigma, E[(X_l)^{\text{LIO}} \texttt{>>=}\ e] \rangle) \\
& = \langle \varepsilon_L(\Sigma), \varepsilon_L(E)[(\varepsilon_L(X_l))^{\text{LIO}} \texttt{>>=}\ \varepsilon_L(e)] \rangle \\
& = \langle \varepsilon_L(\Sigma), \varepsilon_L(E)[(X_l)^{\text{LIO}} \texttt{>>=}\ \varepsilon_L(e)] \rangle \\
\longrightarrow_L\ & \varepsilon_L(\langle \varepsilon_L(\Sigma), \varepsilon_L(E)[(X_l)^{\text{LIO}}] \rangle) \\
& = \langle \varepsilon_L(\Sigma), \varepsilon_L(E)[(\varepsilon_L(X_l))^{\text{LIO}}] \rangle \\
& = \langle \varepsilon_L(\Sigma), \varepsilon_L(E)[\varepsilon_L((X_l)^{\text{LIO}})] \rangle \\
& = \langle \varepsilon_L(\Sigma), \varepsilon_L(E[(X_l)^{\text{LIO}}]) \rangle = \varepsilon_L(\langle \Sigma, E[(X_l)^{\text{LIO}}] \rangle)
\end{aligned}$$

by definition of $\varepsilon_L$ and Proposition 3.

— $l \not\sqsubseteq L$:

$$\begin{aligned}
& \varepsilon_L(\langle \Sigma, E[(X_l)^{\text{LIO}} \mathtt{>>=}\ e]\rangle) \\
& = \langle \varepsilon_L(\Sigma), \varepsilon_L(E)[(\bullet)^{\text{LIO}}]\rangle \\
\longrightarrow_L\ & \varepsilon_L(\langle \varepsilon_L(\Sigma), \varepsilon_L(E)[(\bullet)^{\text{LIO}}]\rangle) \\
& = \langle \varepsilon_L(\Sigma), \varepsilon_L(E)[(\bullet)^{\text{LIO}}]\rangle \\
& = \langle \varepsilon_L(\Sigma), \varepsilon_L(E)[(\varepsilon_L(X_l))^{\text{LIO}}]\rangle \\
& = \langle \varepsilon_L(\Sigma), \varepsilon_L(E)[\varepsilon_L((X_l)^{\text{LIO}})]\rangle \\
& = \langle \varepsilon_L(\Sigma), \varepsilon_L(E[(X_l)^{\text{LIO}}])\rangle = \varepsilon_L(\langle \Sigma, E[(X_l)^{\text{LIO}}]\rangle) \\
& \text{by definition of } \varepsilon_L \text{ and Proposition 3.}
\end{aligned}$$

► Case
$$\frac{\Sigma.\mathtt{lbl} \sqsubseteq l \sqsubseteq \Sigma.\mathtt{clr}}{\langle \Sigma, E[\mathtt{label}\ l\ e]\rangle \longrightarrow \langle \Sigma, E[\mathtt{return}\ (\mathtt{Lb}\ l\ e)]\rangle}:$$

— $l \sqsubseteq L$:

$$\begin{aligned}
& \varepsilon_L(\langle \Sigma, E[\mathtt{label}\ l\ e]\rangle) \\
& = \langle \varepsilon_L(\Sigma), \varepsilon_L(E)[\mathtt{label}\ l\ \varepsilon_L(e)]\rangle \\
\longrightarrow_L\ & \varepsilon_L(\langle \varepsilon_L(\Sigma), \varepsilon_L(E)[\mathtt{return}\ (\mathtt{Lb}\ l\ \varepsilon_L(e))]\rangle) \\
& = \langle \varepsilon_L(\Sigma), \varepsilon_L(E)[\mathtt{return}\ (\mathtt{Lb}\ l\ \varepsilon_L(e))]\rangle \\
& = \langle \varepsilon_L(\Sigma), \varepsilon_L(E)[\varepsilon_L(\mathtt{return}\ (\mathtt{Lb}\ l\ e))]\rangle \\
& = \varepsilon_L(\langle \Sigma, E[\mathtt{return}\ (\mathtt{Lb}\ l\ e)]\rangle)
\end{aligned}$$

— $l \not\sqsubseteq L$:

$$\begin{aligned}
& \varepsilon_L(\langle \Sigma, E[\mathtt{label}\ l\ e]\rangle) \\
& = \langle \varepsilon_L(\Sigma), \varepsilon_L(E)[\mathtt{label}\ l\ \varepsilon_L(e)]\rangle \\
\longrightarrow_L\ & \varepsilon_L(\langle \varepsilon_L(\Sigma), \varepsilon_L(E)[\mathtt{return}\ (\mathtt{Lb}\ l\ \varepsilon_L(e))]\rangle) \\
& = \langle \varepsilon_L(\Sigma), \varepsilon_L(E)[\mathtt{return}\ (\mathtt{Lb}\ l\ \bullet)]\rangle \\
& = \langle \varepsilon_L(\Sigma), \varepsilon_L(E)[\varepsilon_L(\mathtt{return}\ (\mathtt{Lb}\ l\ e))]\rangle \\
& = \varepsilon_L(\langle \Sigma, E[\mathtt{return}\ (\mathtt{Lb}\ l\ e)]\rangle)
\end{aligned}$$

► Case
$$\frac{l' = \Sigma.\mathtt{lbl} \sqcup l \qquad l' \sqsubseteq \Sigma.\mathtt{clr} \qquad \Sigma' = \Sigma[\mathtt{lbl} \mapsto l']}{\langle \Sigma, E[\mathtt{unlabel}\ (\mathtt{Lb}\ l\ e)]\rangle \longrightarrow \langle \Sigma', E[\mathtt{return}\ e]\rangle}:$$

— $l \sqsubseteq L$:

$$\begin{aligned}
&\varepsilon_L(\langle \Sigma, E[\texttt{unlabel}\ (\texttt{Lb}\ l\ e)]\rangle)\\
&= \langle \varepsilon_L(\Sigma), \varepsilon_L(E[\texttt{unlabel}\ (\texttt{Lb}\ l\ e)])\rangle\\
&= \langle \varepsilon_L(\Sigma), \varepsilon_L(E)[\texttt{unlabel}\ (\texttt{Lb}\ l\ \varepsilon_L(e))]\rangle\\
\longrightarrow_L &\ \langle \varepsilon_L(\varepsilon_L(\Sigma^1)), \varepsilon_L(\varepsilon_L(E)[\texttt{return}\ (\varepsilon_L(e))])\rangle\\
&= \langle \varepsilon_L(\Sigma^1), \varepsilon_L(E)[\texttt{return}\ \varepsilon_L(e)]\rangle\\
&= \langle \varepsilon_L(\Sigma^1), \varepsilon_L(E)[\varepsilon_L(\texttt{return}\ e)]\rangle\\
&= \langle \varepsilon_L(\Sigma^1), \varepsilon_L(E[\texttt{return}\ e])\rangle\\
&= \langle \varepsilon_L(\Sigma'), \varepsilon_L(E[\texttt{return}\ e])\rangle\\
&= \varepsilon_L(\langle \Sigma', E[\texttt{return}\ e]\rangle)
\end{aligned}$$

where $\varepsilon_L(\Sigma^1) = \varepsilon_L(\Sigma[\texttt{lbl} \mapsto l'])$, and thus it directly follows that $\varepsilon_L(\Sigma^1) = \varepsilon_L(\Sigma')$.

— $l \not\sqsubseteq L$:

$$\begin{aligned}
&\varepsilon_L(\langle \Sigma, E[\texttt{unlabel}\ (\texttt{Lb}\ l\ e)]\rangle)\\
&= \langle \varepsilon_L(\Sigma), \varepsilon_L(E[\texttt{unlabel}\ (\texttt{Lb}\ l\ e)])\rangle\\
&= \langle \varepsilon_L(\Sigma), \varepsilon_L(E)[\texttt{unlabel}\ (\texttt{Lb}\ l\ \bullet)]\rangle\\
\longrightarrow_L &\ \varepsilon_L(\langle \varepsilon_L(\Sigma^1), \varepsilon_L(E)[\texttt{return}\ \bullet]\rangle)\\
&= \langle \varepsilon_L(\varepsilon_L(\Sigma^1)), \bullet\rangle\\
&= \varepsilon_L(\langle \Sigma', E[\texttt{return}\ e]\rangle)
\end{aligned}$$

The last steps holds, as in the second case of `return`, because $\Sigma'.\texttt{lbl} \not\sqsubseteq L$ and any term is erased to $\bullet$. Similarly, $\varepsilon_L(\Sigma^1) = \varepsilon_L(\Sigma')$ follows as before.

▶ We show an example case of the "violating rules":

$$\frac{l \not\sqsubseteq \Sigma.\texttt{clr}}{\langle \Sigma, E[\texttt{unlabel}\ (\texttt{Lb}\ l\ e)]\rangle \longrightarrow \langle \Sigma, E[\texttt{throwLIO}\ X]\rangle}:$$

— $l \sqsubseteq L$:

$$\begin{aligned}
&\varepsilon_L(\langle \Sigma, E[\texttt{unlabel}\ (\texttt{Lb}\ l\ e)]\rangle)\\
&= \langle \varepsilon_L(\Sigma), \varepsilon_L(E[\texttt{unlabel}\ (\texttt{Lb}\ l\ e)])\rangle\\
&= \langle \varepsilon_L(\Sigma), \varepsilon_L(E)[\texttt{unlabel}\ (\texttt{Lb}\ l\ \varepsilon_L(e))]\rangle\\
\longrightarrow_L &\ \langle \varepsilon_L(\varepsilon_L(\Sigma)), \varepsilon_L(\varepsilon_L(E)[\texttt{throwLIO}\ X])\rangle\\
&\langle \varepsilon_L(\Sigma), \varepsilon_L(E)[\texttt{throwLIO}\ \varepsilon_L(X)]\rangle\\
&\langle \varepsilon_L(\Sigma), \varepsilon_L(E)[\texttt{throwLIO}\ X]\rangle\\
&\langle \varepsilon_L(\Sigma), \varepsilon_L(E)[\varepsilon_L(\texttt{throwLIO}\ X)]\rangle\\
&= \varepsilon_L(\langle \Sigma, E[\texttt{throwLIO}\ X]\rangle)
\end{aligned}$$

— $l \not\sqsubseteq L$:

$$
\begin{aligned}
&\varepsilon_L(\langle \Sigma, E[\texttt{unlabel}\ (\texttt{Lb}\ l\ e)]\rangle)\\
&= \langle \varepsilon_L(\Sigma), \varepsilon_L(E[\texttt{unlabel}\ (\texttt{Lb}\ l\ e)])\rangle\\
&= \langle \varepsilon_L(\Sigma), \varepsilon_L(E)[\texttt{unlabel}\ (\texttt{Lb}\ l\ \bullet)]\rangle\\
\longrightarrow_L\ &\langle \varepsilon_L(\varepsilon_L(\Sigma)), \varepsilon_L(\varepsilon_L(E)[\texttt{throwLIO}\ X])\rangle\\
&\langle \varepsilon_L(\Sigma), \varepsilon_L(E)[\texttt{throwLIO}\ \varepsilon_L(X)]\rangle\\
&\langle \varepsilon_L(\Sigma), \varepsilon_L(E)[\texttt{throwLIO}\ X]\rangle\\
&\langle \varepsilon_L(\Sigma), \varepsilon_L(E)[\varepsilon_L(\texttt{throwLIO}\ X)]\rangle\\
&= \varepsilon_L(\langle \Sigma, E[\texttt{throwLIO}\ X]\rangle)
\end{aligned}
$$

▶ Case $\dfrac{\Sigma.\texttt{lbl} \sqsubseteq l \sqsubseteq \Sigma.\texttt{clr} \qquad \Sigma' = \Sigma.\phi[a \mapsto \texttt{Lb}\ l\ e]}{\langle \Sigma, E[\texttt{newLIORef}\ l\ e]\rangle \longrightarrow \langle \Sigma', E[\texttt{return}\ a]\rangle} a\ \textit{fresh}$:

— $l \sqsubseteq L$:

$$
\begin{aligned}
&\varepsilon_L(\langle \Sigma, E[\texttt{newLIORef}\ l\ e]\rangle)\\
&= \langle \varepsilon_L(\Sigma), \varepsilon_L(E[\texttt{newLIORef}\ l\ e])\rangle\\
&= \langle \varepsilon_L(\Sigma), \varepsilon_L(E)[\texttt{newLIORef}\ l\ \varepsilon_L(e)]\\
\longrightarrow_L\ &\langle \varepsilon_L(\varepsilon_L(\Sigma^1)), \varepsilon_L(\varepsilon_L(E)[\texttt{return}\ \varepsilon_L(a)])\rangle\\
&= \langle \varepsilon_L(\Sigma^1), \varepsilon_L(E[\texttt{return}\ a])\rangle\\
&= \varepsilon_L(\langle \Sigma', E[\texttt{return}\ a]\rangle)\rangle,
\end{aligned}
$$

where $\varepsilon_L(\Sigma^1) = \varepsilon_L(\Sigma).\phi[a \mapsto \texttt{Lb}\ l\ e]$, and so $\varepsilon_L(\Sigma^1) = \varepsilon_L(\Sigma')$ follows directly.

— $l \not\sqsubseteq L$: as above. However, in this case, $\varepsilon_L(\Sigma^1) = \varepsilon_L(\Sigma).\phi[a \mapsto \texttt{Lb}\ l\ \bullet]$. From $\varepsilon_L(\Sigma^1).\phi(a) = \varepsilon_L(\texttt{Lb}\ l\ \bullet) = \varepsilon_L(\texttt{Lb}\ l\ e) = \varepsilon_L(\Sigma').\phi(a)$ it follows that $\varepsilon_L(\Sigma^1) = \varepsilon_L(\Sigma')$.

▶ Case
$\dfrac{l' = \Sigma.\texttt{lbl} \sqcup l \qquad l' \sqsubseteq \Sigma.\texttt{clr} \qquad \Sigma' = \Sigma[\texttt{lbl} \mapsto l']}{\langle \Sigma, E[\texttt{catch}\ (X_l)^{\text{LIO}}\ e]\rangle \longrightarrow \langle \Sigma', E[e\ X]\rangle}$:

— $l \sqsubseteq L$:

$$
\begin{aligned}
&\varepsilon_L(\langle \Sigma, E[\texttt{catch}\ (X_l)^{\text{LIO}}\ e]\rangle)\\
&= \langle \varepsilon_L(\Sigma), \varepsilon_L(E[\texttt{catch}\ (X_l)^{\text{LIO}}\ \varepsilon_L(e)])\rangle\\
&= \langle \varepsilon_L(\Sigma), \varepsilon_L(E)[\texttt{catch}\ (X_l)^{\text{LIO}}\ \varepsilon_L(e)]\rangle\\
\longrightarrow_L\ &\langle \varepsilon_L(\varepsilon_L(\Sigma^1)), \varepsilon_L(\varepsilon_L(E)[\varepsilon_L(e)\ X])\rangle\\
&= \langle \varepsilon_L(\Sigma^1), \varepsilon_L(E)[\varepsilon_L(e)\ \varepsilon_L(X)]\rangle\\
&= \langle \varepsilon_L(\Sigma^1), \varepsilon_L(E)[\varepsilon_L(e)\ X]\rangle\\
&= \langle \varepsilon_L(\Sigma^1), \varepsilon_L(E)[\varepsilon_L(e\ X)]\rangle\\
&= \langle \varepsilon_L(\Sigma^1), \varepsilon_L(E[e\ X])\rangle\\
&= \langle \varepsilon_L(\Sigma'), \varepsilon_L(E[e\ X])\rangle\\
&= \varepsilon_L(\langle \Sigma', E[e\ X]\rangle)
\end{aligned}
$$

where $\varepsilon_L(\Sigma^1) = \varepsilon_L(\Sigma[\mathtt{lbl} \mapsto l'])$, and directly $\varepsilon_L(\Sigma^1) = \varepsilon_L(\Sigma')$.

— $l \not\sqsubseteq L$:

$$\begin{aligned}
& \varepsilon_L(\langle \Sigma, E[\mathtt{catch}\ (X_l)^{\text{LIO}}\ e] \rangle) \\
& = \langle \varepsilon_L(\Sigma), \varepsilon_L(E[\mathtt{catch}\ (X_l)^{\text{LIO}}\ \varepsilon_L(e)]) \rangle \\
& = \langle \varepsilon_L(\Sigma), \varepsilon_L(E)[\mathtt{catch}\ (\bullet)^{\text{LIO}} \varepsilon_L(e)\ ] \rangle \\
\longrightarrow_L\ & \varepsilon_L(\langle \varepsilon_L(\Sigma^1), \varepsilon_L(E)[\varepsilon_L(e)\ \bullet] \rangle) \\
& = \langle \varepsilon_L(\Sigma^1), \bullet \rangle \\
& = \varepsilon_L(\langle \Sigma', E[e\ X] \rangle)
\end{aligned}$$

The last steps holds since $\Sigma^1.\mathtt{lbl} \not\sqsubseteq L$ and any term is erased to $\bullet$. As before, $\varepsilon_L(\Sigma^1) = \varepsilon_L(\Sigma')$ trivialy holds.

□

*Lemma 3* (*Simulation*)

If $\Gamma \vdash e : \tau$ and $\langle \Sigma, e \rangle \longrightarrow^* \langle \Sigma', e' \rangle$ then $\varepsilon_L(\langle \Sigma, e \rangle) \longrightarrow_L^* \varepsilon_L(\langle \Sigma', e' \rangle)$.

*Proof*

Lemma 2 shows the multi-step simulation for expressions that do not execute `toLabeled`. Thus, to show the general multi-step simulation we must first show that `toLabeled` preserves the simulation. The general simulation follows directly.

The proof for the simulation of `toLabeled` follows by induction on the number of executed `toLabeled`. The base case consists of a single `toLabeled`. Specifically, for a computation with a single executed `toLabeled`, we have:

$$\langle \Sigma, e \rangle \longrightarrow^* \langle \Sigma', e' \rangle,$$

that can be expanded into

$$\langle \Sigma, e \rangle \longrightarrow^* \langle \Sigma_0, E[\mathtt{toLabeled}\ l\ e_0] \rangle \longrightarrow \langle \Sigma_0'', E[\mathtt{label}\ l\ v] \rangle \longrightarrow^* \langle \Sigma', e' \rangle,$$

where $\langle \Sigma_0, e_0 \rangle \longrightarrow^* \langle \Sigma_0', (v)^{\text{LIO}} \rangle$, and $\Sigma_0'' = \Sigma_0'[\mathtt{lbl} \mapsto \Sigma_0.\mathtt{lbl}, \mathtt{clr} \mapsto \Sigma_0.\mathtt{clr}]$. The expansion highlights the first occurrence of a `toLabeled`, and so $e_0$, and $e'$ do not have any additional `toLabeled`s. From this observation it is clear that the simulation of the base case follows directly by Lemma 2. Specifically, to show the simulation for (TOLAB-1)

$$\frac{\langle \Sigma_0, e_0 \rangle \longrightarrow^* \langle \Sigma_0', (v)^{\text{LIO}} \rangle \qquad \begin{array}{c}\Sigma_0.\mathtt{lbl} \sqsubseteq l \sqsubseteq \Sigma_0.\mathtt{clr} \\ \Sigma_0'.\mathtt{lbl} \sqsubseteq l \end{array} \qquad \Sigma_0'' = \Sigma_0'[\mathtt{lbl} \mapsto \Sigma_0.\mathtt{lbl}, \mathtt{clr} \mapsto \Sigma_0.\mathtt{clr}]}{\langle \Sigma_0, E[\mathtt{toLabeled}\ l\ e_0] \rangle \longrightarrow \langle \Sigma_0'', E[\mathtt{label}\ l\ v] \rangle},$$

where $e_0$ does not have any `toLabeled` we need only show the simulation of the conclusion; the simulation of the big step in the premise follows directly from Lemma 2. We show this below:

$$\begin{aligned} &\varepsilon_L(\langle \Sigma_0, E[\texttt{toLabeled}\ l\ (e_0)^{\text{LIO}}]\rangle) \\ &= \langle \varepsilon_L(\Sigma_0), \varepsilon_L(E[\texttt{toLabeled}\ l\ (e_0)^{\text{LIO}}])\rangle \\ &= \langle \varepsilon_L(\Sigma_0), \varepsilon_L(E)[\texttt{toLabeled}\ l\ (\varepsilon_L(e_0))^{\text{LIO}}]\rangle \\ \longrightarrow_L\ &\langle \varepsilon_L(\varepsilon_L(\Sigma_0'')), \varepsilon_L(\varepsilon_L(E)[\texttt{label}\ l\ \varepsilon_L(v)])\rangle \\ &= \langle \varepsilon_L(\Sigma_0''), \varepsilon_L(E)[\texttt{label}\ l\ \varepsilon_L(v)]\rangle \\ &= \langle \varepsilon_L(\Sigma_0''), \varepsilon_L(E[\texttt{label}\ l\ v])\rangle \\ &= \varepsilon_L(\langle \Sigma_0'', E[\texttt{label}\ l\ v]\rangle) \end{aligned}$$

Correspondingly, the simulation of the $\langle \Sigma_0'', E[\texttt{label}\ l\ v]\rangle \longrightarrow^* \langle \Sigma', e'\rangle$ step follows directly by Lemma 2. The simulations of (TOLAB-2) follows similarly.

It is worth noting that the simulation of (BIND-1), as proved in Lemma 1, holds for exception labels, irrespective of the current label. This is a necessary condition when a computation executes `toLabeled` as the current label and exception label may not always be the same.

Our inductive hypothesis states that the simulation of

$$\langle \Sigma, e\rangle \longrightarrow^* \langle \Sigma', e'\rangle,$$

holds for the case where `toLabeled` is executed $k$ times. With this assumption, the simulation of

$$\langle \Sigma, e\rangle \longrightarrow^* \langle \Sigma', e'\rangle,$$

with $k+1$ `toLabeled` executions, follows in a similar manner to the base case. Specifically, searching for the first `toLabeled` and expanding, we have:

$$\overbrace{\langle \Sigma, e\rangle \longrightarrow^* \langle \Sigma_0, E}^{\text{first big-step}}\overbrace{[\texttt{toLabeled}\ l\ e_0]\rangle \longrightarrow^* \langle \Sigma', e'\rangle}^{\text{second big-step}}$$

where at most $k$ `toLabeled`s could have been executed in the first big-step, the inner computation $e_0$, or the second big-step. The simulation of all these execution steps follows by application of the inductive hypothesis. □

*Theorem 1* (*Non-interference*)
Given a computation $e$ (with no $\bullet$, $(\ )^{\text{LIO}}$, `Lb`, or $X_l$) where $\Gamma \vdash e : \texttt{Labeled}\ \ell\ \tau \to \texttt{LIO}\ \ell\ (\texttt{Labeled}\ \ell\ \tau')$, environments $\Sigma_1$ and $\Sigma_2$ where $\Sigma_1.\phi = \Sigma_2.\phi = \emptyset$, security label $l$, an attacker at level $L$ such that $l \sqsubseteq L$, then

$$\begin{aligned} &\forall e_1 e_2.(\Gamma \vdash e_i : \texttt{Labeled}\ \ell\ \tau)_{i=1,2} \wedge (e_i = \texttt{Lb}\ l\ e_i')_{i=1,2} \wedge \langle \Sigma_1, e_1\rangle \approx_L \langle \Sigma_2, e_2\rangle \\ &\quad \wedge \langle \Sigma_1, e\ e_1\rangle \longrightarrow^* \langle \Sigma_1', (v_1)^{\text{LIO}}\rangle \wedge \langle \Sigma_2, e\ e_2\rangle \longrightarrow^* \langle \Sigma_2', (v_2)^{\text{LIO}}\rangle \\ &\quad \Longrightarrow \langle \Sigma_1', (v_1)^{\text{LIO}}\rangle \approx_L \langle \Sigma_2', (v_2)^{\text{LIO}}\rangle \end{aligned}$$

*Proof*
From Lemma 3, for $i = 1,2$, we have

$$\varepsilon_L(\langle \Sigma_i, e\ (\texttt{Lb}\ l\ e_i')\rangle) \longrightarrow_L^* \varepsilon_L(\langle \Sigma_i', (v_i)^{\text{LIO}}\rangle),$$

where $v_i = \texttt{Lb}\ l_i\ e_i''$ or $v_i = X_{l''}$. First we highlight that:

$$\varepsilon_L(\langle \Sigma, e\rangle) = \varepsilon_L(\langle \Sigma', e'\rangle) \implies \langle \Sigma, e\rangle \approx_L \langle \Sigma', e'\rangle$$

Note that the converse is not necessarily true, since the stores may differ in the references with labels above $L$. Then, from the determinacy of $\longrightarrow_L$, given in Proposition 2, and since the starting environment configurations are the same (observe that $\langle\Sigma_1, e\ (\texttt{Lb}\ l\ e'_1)\rangle \approx_L \langle\Sigma_2, e\ (\texttt{Lb}\ l\ e'_2)\rangle \implies \varepsilon_L(\langle\Sigma_1, e\ (\texttt{Lb}\ l\ e'_1)\rangle) = \varepsilon_L(\langle\Sigma_2, e\ (\texttt{Lb}\ l\ e'_2)\rangle)$ since $\Sigma_1.\phi = \Sigma_2.\phi = \emptyset$), it must be that the end environment configurations also be the same, i.e., $\varepsilon_L(\langle\Sigma'_1, (v_1)^{\text{LIO}}\rangle) = \varepsilon_L(\langle\Sigma'_2, (v_2)^{\text{LIO}}\rangle)$. The $L$-equivalence directly follows from the above observation.

For completeness, we detail the following cases:

- Case $\Sigma_i.\texttt{lbl} \not\sqsubseteq L$: We have

$$\varepsilon_L(\langle\Sigma_i, e\ (\texttt{Lb}\ l\ e'_i)\rangle) = \langle\varepsilon_L(\Sigma_i), \bullet\rangle \longrightarrow^*_L \langle\varepsilon_L(\Sigma'_i), \bullet\rangle = \varepsilon_L(\langle\Sigma'_i, (v_i)^{\text{LIO}}\rangle).$$

  From the determinacy of $\longrightarrow_L$, it must be that the end environment configurations are the same, from which it directly follows that $\langle\Sigma'_1, (v_1)^{\text{LIO}}\rangle \approx_L \langle\Sigma'_2, (v_2)^{\text{LIO}}\rangle$.
- Case $\Sigma_i.\texttt{lbl} \sqsubseteq L \wedge \Sigma'_i.\texttt{lbl} \not\sqsubseteq L$: We have

$$\varepsilon_L(\langle\Sigma_i, e\ (\texttt{Lb}\ l\ e'_i)\rangle) = \langle\varepsilon_L(\Sigma_i), \varepsilon_L(e)\ (\texttt{Lb}\ l\ \varepsilon_L(e'_i))\rangle \longrightarrow^*_L \langle\varepsilon_L(\Sigma'_i), \bullet\rangle = \varepsilon_L(\langle\Sigma'_i, (v_i)^{\text{LIO}}\rangle).$$

  As before, since the initial environment configurations are the same, from the determinacy of $\longrightarrow_L$ we end with the same configuration, which directly corresponds to $L$-equivalence.
- Case $\Sigma_i.\texttt{lbl} \sqsubseteq L \wedge \Sigma'_i.\texttt{lbl} \sqsubseteq L$: We have

$$\varepsilon_L(\langle\Sigma_i, e\ (\texttt{Lb}\ l\ e'_i)\rangle) = \langle\varepsilon_L(\Sigma_i), \varepsilon_L(e)\ (\texttt{Lb}\ l\ \varepsilon_L(e'_i))\rangle \longrightarrow^*_L \langle\varepsilon_L(\Sigma'_i), (v_i)^{\text{LIO}}\rangle.$$

  From the determinacy of $\longrightarrow_L$, it must be that $\varepsilon_L(\Sigma'_1) = \varepsilon_L(\Sigma'_2)$ and $\varepsilon_L(v_1) = \varepsilon_L(v_2)$, and directly $\langle\Sigma'_1, (v_1)^{\text{LIO}}\rangle \approx_L \langle\Sigma'_2, (v_2)^{\text{LIO}}\rangle$.

□

*Theorem 2* (*Store confinement*)
Given labels $l$ and $l_c$, a computation $e$ (with no $\bullet$, $a$, $(\ )^{\text{LIO}}$, $\texttt{Lb}$, or $X_{l'}$) such that $\Gamma \vdash e : \texttt{LIO}\ \ell\ \tau$, and environment $\Sigma[\texttt{lbl} \mapsto l, \texttt{clr} \mapsto l_c]$ where $l \sqsubseteq l_c$, then

$$\langle\Sigma, e\rangle \longrightarrow^* \langle\Sigma', (v)^{\text{LIO}}\rangle \implies (\Sigma.\phi)_{\downarrow l} = (\Sigma'.\phi)_{\downarrow l} \wedge (\Sigma.\phi)_{\uparrow l_c} = (\Sigma'.\phi)_{\uparrow l_c}$$

*Proof*
By contradiction. We show the case of creating new referneces, the case of modifying an existing reference follows similarly. Suppose that

- $(\Sigma.\phi)_{\downarrow l} \neq (\Sigma'.\phi)_{\downarrow l}$. Then, $\exists(a, \texttt{Lb}\ l_v\ e_a) \in (\Sigma'.\phi)_{\downarrow l}.(a, \texttt{Lb}\ l_v\ e_a) \notin (\Sigma.\phi)_{\downarrow l}$ and $l \not\sqsubseteq l_v$. Moreover, (by Proposition 4) there must be a step at which point the new reference is created: $\langle\Sigma', e\rangle \longrightarrow^* \langle\Sigma_a, E[\texttt{newLIORef}\ l_v\ e_a]\rangle \longrightarrow^* \langle\Sigma', (v)^{\text{LIO}}\rangle$, such that $(\Sigma.\phi)_{\downarrow l} = (\Sigma_a.\phi)_{\downarrow l}$. However, by (NREF) it must be that $l \sqsubseteq l_v$. Hence, we have a contradiction.
- $(\Sigma.\phi)_{\uparrow l_c} \neq (\Sigma'.\phi)_{\uparrow l_c}$. Then, $\exists(a, \texttt{Lb}\ l_v\ e_a) \in (\Sigma'.\phi)_{\uparrow l_c}.(a, \texttt{Lb}\ l_v\ e_a) \notin (\Sigma.\phi)_{\uparrow l_c}$ and $l_v \not\sqsubseteq l_c$. Moreover, (by Proposition 5) there must be a step at which point the new reference is created: $\langle\Sigma', e\rangle \longrightarrow^* \langle\Sigma_a, E[\texttt{newLIORef}\ l_v\ e_a]\rangle \longrightarrow^* \langle\Sigma', (v)^{\text{LIO}}\rangle$, such that $(\Sigma.\phi)_{\uparrow l_c} = (\Sigma_a.\phi)_{\uparrow l_c}$. However, by (NREF) it must be that $l_v \sqsubseteq l_c$. Hence, we have a contradiction.

□

*Theorem 3* (*`Labeled` creation confinement*)

Given labels $l$, $l_c$, and $l_v$, a computation $e$ (with no $\bullet$, $a$, $(\,)^{\text{LIO}}$, $\texttt{Lb}$, or $X_{l'}$) where $\Gamma \vdash e : \texttt{LIO}\ \ell\ (\texttt{Labeled}\ \ell\ \tau)$, and environment $\Sigma[\texttt{lbl} \mapsto l, \texttt{clr} \mapsto l_c]$ such that $l \sqsubseteq l_c$, then

$$\langle \Sigma, e \rangle \longrightarrow^* \langle \Sigma', (\texttt{Lb}\ l_v\ e_1)^{\text{LIO}} \rangle \implies l \sqsubseteq l_v \sqsubseteq l_c \vee \exists (a, \texttt{Lb}\ l_1\ e_1') \in \Sigma.\phi . \texttt{Lb}\ l_v\ e_1\ \tilde{\in}\ e_1' \wedge l_1 \sqsubseteq l_c$$

Here, operator $\tilde{\in}$ is defined as the syntactic appearance of the left-hand expression into the right-hand side operand.

*Proof*

Since $e$ cannot contain $\texttt{Lb}$s, the final labeled value must be created or retrieved from the store. By induction on expressions and evaluation contexts and using Propositions 4 and 5 it must be that the label of the value is bounded by the initial current label and clearance or the labeled value appears syntactically in the store. The proof follows by case analysis on how a $\texttt{Labeled}$ value can be obtained.

Suppose the value is created, i.e., $\nexists (a, \texttt{Lb}\ l_1\ e_1') \in \Sigma.\phi . \texttt{Lb}\ l_v\ e_1\ \tilde{\in}\ e_1' \wedge l_1 \sqsubseteq l_c$ Then, there must be an intermediate step where the labeled value is created. Specifically, $\langle \Sigma, e \rangle \longrightarrow^* \langle \Sigma_1, E[\texttt{label}\ l_v\ e_1] \rangle \longrightarrow^* \langle \Sigma', (\texttt{Lb}\ l_v\ e_1)^{\text{LIO}} \rangle$. (Recall that $\texttt{toLabeled}$ also reduces to $\texttt{label}$, thus we need only handle this case.) Hence, from rule (LAB) it must be that $\Sigma_1.\texttt{lbl} \sqsubseteq l_v \sqsubseteq \Sigma_1.\texttt{clr}$ and by Propositions 4 and 5 it must be that $l \sqsubseteq l_v \sqsubseteq l_c$.

Suppose that the value is not created with $\texttt{label}$. Then, $\exists (a, \texttt{Lb}\ l_1\ e_1') \in \Sigma.\phi . \texttt{Lb}\ l_v\ e_1\ \tilde{\in}\ e_1'$ and $l_1 \sqsubseteq l_c$. The $l_1 \sqsubseteq l_c$ must hold since there must be an intermediate step where the reference is read. Specifically, $\langle \Sigma, e \rangle \longrightarrow^* \langle \Sigma_1, E[\texttt{readLIORef}\ a] \rangle \longrightarrow \langle \Sigma_1', E[\texttt{return}\ e_1'] \rangle \longrightarrow^* \langle \Sigma', (\texttt{Lb}\ l_v\ e_1)^{\text{LIO}} \rangle$. From rule (RREF) it must be that $l_1 \sqsubseteq \Sigma_1.\texttt{clr}$ and by Proposition 5 it directly follows that $l_1 \sqsubseteq l_c$. Because our semantics does not have the evaluation context $E ::= \cdots \mid \texttt{Lb}\ e\ E$, the values of references are not always evaluated and thus the labeled value $\texttt{Lb}\ l_v\ e_1$ must syntactically appear in $e_1'$. For example, if $e_1 = 2$, it holds that $e_1\ \tilde{\in}\ \lambda x.\texttt{if}\ x\ \texttt{then}\ (\texttt{Lb}\ l_v\ 2)\ \texttt{else}\ (\texttt{Lb}\ l_w\ 3)$, but not $e_1\ \tilde{\in}\ \lambda x.\texttt{if}\ x\ \texttt{then}\ (\texttt{Lb}\ l_v\ (1+1))\ \texttt{else}\ (\texttt{Lb}\ l_w\ 3)$ for some $l_w$. $\square$